\documentclass[10pt]{iopart}
\usepackage{graphicx,amsfonts,amssymb}

\newcommand{\wb}{\omega_{\mathrm{b}}}
\newcommand{\wc}{\omega_{\mathrm{b,c}}}
\newcommand{\Tb}{T_{\mathrm{b}}}

\newcommand{\wt}{\wb\, t}
\newcommand{\sgn}{\mathrm{sign}}
\newcommand{\ii}{\mathrm{i}}
\newcommand{\ee}{\varepsilon}

\newcommand{\sign}{\mathrm{sign}}

\newcommand{\fracc}[2]{\frac{\displaystyle #1}{\displaystyle #2}}

\newcommand{\middlefig}{.45\textwidth}

\newcommand{\F}{\mathcal{F}}

\renewcommand{\a}{\alpha}
\renewcommand{\t}{\vartheta}
\newcommand{\la}{\lambda}

\renewcommand{\k}{\kappa}

\newcommand{\cn}{\mathrm{cn}}

\newcommand{\sn}{\mathrm{sn}}
\newcommand{\dn}{\mathrm{dn}}
\renewcommand{\ns}{\mathrm{ns}}
\newcommand{\ds}{\mathrm{ds}}
\newcommand{\cs}{\mathrm{cs}}

\newcommand{\E}{\mathcal{E}}

\newcommand{\dd}{\mathrm{d}}
\newcommand{\J}{{\mathcal J}}

\begin{document}

\title{Stability of non-time-reversible phonobreathers}
\author{J Cuevas$^1$, JFR Archilla$^2$ and FR Romero$^3$}

\address{$^1$ Grupo de F\'{\i}sica No Lineal. Departamento de F\'\i sica Aplicada I, Escuela Polit\'ecnica Superior.
Universidad de Sevilla. Virgen de \'Africa 7, 41011 Sevilla, Spain}

\address{$^2$ Grupo de F\'{\i}sica No Lineal. Departamento de F\'\i sica Aplicada I,
Escuela T\'ecnica Superior de Ingenier\'{\i}a Inform\'atica.
Universidad de Sevilla. Avda. Reina Mercedes, s/n, 41012 Sevilla, Spain}

\address{$^3$ Grupo de F\'{\i}sica No Lineal. \'Area de F\'\i sica Te\'orica,
Facultad de F\'{\i}sica.
Universidad de Sevilla. Avda. Reina Mercedes, s/n, 41012 Sevilla, Spain}

\ead{jcuevas@us.es}

\begin{abstract}

Non-time reversible phonobreathers are non-linear waves that can transport energy in coupled oscillator chains by means of a phase-torsion mechanism. In this paper, the stability properties of these structures have been considered. It has been performed an analytical study for low-coupling solutions based upon the so called {\em multibreather stability theorem} previously developed by some of the authors [Physica D {\bf 180} 235]. A numerical analysis confirms the analytical predictions and gives a detailed picture of the existence and stability properties for arbitrary frequency and coupling.
\end{abstract}

\pacs{63.20.Ry}
\submitto{\JPA}
\maketitle

\section{Introduction}
\label{sec:introduction}

One of the subjects where a great deal of attention has been focused in the last two decades is the dynamics of nonlinear lattices. Intrinsic localized modes or {\em discrete breathers} is one of the most outstanding structures that arise in those lattices \cite{Aubry06,FG08}. Discrete breathers are periodic and localized solutions whose existence is allowed by the interplay between discreteness and nonlinearity. The existence of those
structures in networks of anharmonic oscillators (also known as Klein-Gordon lattices) was firstly proven at 1994 by MacKay and Aubry \cite{MA94}. They established that discrete breathers can be continued from the anti-continuous limit (i.e. the limit where there is no coupling between the oscillators) to finite coupling as long as no integer multiple of the breather frequency resonates with the linear modes in the phonon band.

This theorem also demonstrates the existence of breathers with more than one excited sites, dubbed as {\em multibreathers}. These multibreathers receive special names in some cases; for instance, breathers with all their sites excited receive the name of {\em phonobreathers}. If all but one of the sites are excited the multibreather is called a {\em dark breather} \cite{AACR02}, in analogy to the dark solitons existing in the Discrete Nonlinear Schr\"odinger (DNLS) equation \cite{MJKA02,JK99}.

In order to observe experimentally discrete breathers, they have to be stable. The stability of one-site breathers (i.e. breathers with only one excited site) was firstly proven in \cite{A97,MS98}. It was not until 2003 where some of the authors of the present paper developed a method for determining the stability or instability of many kinds of multibreathers \cite{ACSA03,CAR05} based on the Aubry's band theory developed in Ref.~\cite{A97}. An alternative approach, introduced in \cite{KK09}, is based in previous work by MacKay {\em et al} \cite{AMS01,MS02,mackay}.
Recently, the equivalence between both approaches has been proven \cite{Arxiv}.

Most of theoretical work related to discrete breathers has been carried on time-reversible solutions of Hamiltonian lattices. Non-time-reversible solutions have mostly been considered in dissipative lattices \cite{FG08b,CCK09,MFMF01,MMFF03}, which have many experimental applications in chains of coupled pendula \cite{CEKA09}, micromechanical arrays \cite{SHS06}, transmission lines \cite{EBS08,EPSKB10} and Josephson junction arrays
\cite{TMO00,BAUFZ00}.

However, there have been very few approaches to non-time-reversible breathers or multibreathers in Hamiltonian lattices. The existence of one of such solutions was firstly proven in the pioneer work of MacKay and Aubry \cite{MA94}. It was restricted to phonobreathers where there is a constant phase
difference between adjacent sites. These solutions can transport energy by means of {\em phase torsion} and can be viewed as nonlinear phonons or {\em phasons}. Later on, Aubry \cite{A97} demonstrated the existence of a generic non-time-reversible breather, independent on the excited sites. The proof was also extended to inhomogeneous lattices (e.g. with vacancies) and vortices. It was also proven that phonobreathers with phase torsion generate a stationary flux, and that time-reversible phonobreathers cannot transport energy. Finally, Cretegny and Aubry numerically demonstrate \cite{CA97,CA98} the existence of non-time-reversible phonobreathers in homogenous and inhomogeneous (with vacancies) 1-D lattices and vortices and breather ``rivers''
(i.e. percolating clusters of breathers connecting two boundaries submitted to phase torsion) in 2D square lattices. They also sketch some properties of the flux and its stability in 1D lattices. Further applications of this theory can be found in many publications on discrete vortices in DNLS
\cite{MK01,CMK07,LKKKFB08,CJKL09,CHSK09} and Klein-Gordon \cite{Arxiv,KM05} lattices.

The aim of this paper is twofold. On the one hand, we will prove by making use of the techniques developed in \cite{ACSA03,CAR05}, a result mentioned in passing on Ref.~\cite{CA97}: the stability of phonobreathers depends on the phase difference between adjacent sites and the phonobreather frequency. On the other hand, we will make an analysis of the stability and flux dependence on the system parameters for two kind of potentials in Klein-Gordon
lattices, checking the validity of the analytical predictions.

The paper is organized as follows: in Section \ref{sec:model} we introduce the model equations; section \ref{sec:MST} deals with the analytical results regarding the stability for low coupling; in order to check these results, we make an exhaustive numerical analysis of the existence and stability of phonobreathers at finite coupling in Section \ref{sec:stab}; finally, we present our conclusions and some possible extension of this work in Section
\ref{sec:conc}.

\section{Model setup}
\label{sec:model}

\subsection{Dynamical equations and energy flux}

We a consider a Klein--Gordon chain of oscillators with nearest-neighbours harmonic coupling. The dynamical equations are of the form:

\begin{equation}\label{eq:dyn}
    \ddot{u}_n \,+\,V'(u_n)\,+\,\ee (2u_n-u_{n+1}-u_{-1})\,=\,0\,\quad n=1,\dots,N \label{eq:klein}
\end{equation}
where the variables $u_n$ are the displacements with respect to the equilibrium positions, $V(u_n)$ is the on--site potential, $N$ is the number of oscillators, and $\ee>0$ is the coupling constant.

We consider two paradigmatic cases of on-site potentials: Morse (soft) potential

\begin{equation}
    V(u)=\frac{1}{2}[\exp(-u)-1]^2
\end{equation}
and $\phi^4$ (quartic) hard potential

\begin{equation}
    V(u)=\frac{1}{2}u^2+\frac{1}{4}u^4
\end{equation}

We look for non-time-reversible solutions of the dynamical equations (\ref{eq:dyn}) with all the particles excited, so that the difference between the phases of the temporal oscillations of two nearest-neighbours lattice sites is a constant $\alpha$. Thus, there is a phase torsion between the boundaries of the lattice of value $\tau=N\alpha$. In order to fit the periodic boundary conditions, $\tau=2\pi m$ with $m\in\mathbb{Z}$, and, in consequence, $\alpha=2\pi m/N$. Hence, we are dealing with an anharmonic plane wave (or a phonobreather) with a wave number equal to $\alpha$ that transmits an energy flux along the chain of
oscillators \cite{CA97}.

Let us remark the similarities of non-time-reversible phonobreathers with q-breathers, which are localized solutions in the reciprocal space (see \cite{QB1} for FPU and \cite{QB2} for DNLS lattices). q-breathers can be generally considered as the superposition of a finite number of normal modes. The solutions considered in the present paper, i.e. phonobreathers, are also localized in the reciprocal space, so they can be cast as a special case of q-breathers, consisting only of a normal mode characterized by the wavenumber $\alpha$.

In order to calculate phonobreather solutions we make use of methods based on the anti-continuous limit~\cite{MA94,MA96}, that is, an orbit of frequency $\wb$ for each isolated oscillator is calculated and the coupling constant is subsequently varied with a path-following (Newton-Raphson) method. In this paper, we have used a Fourier space implementation of the dynamical equations.

Fourier space methods are based on the fact that the solutions are $\Tb$-periodic (For a detailed explanation of these methods, the reader is referred to Refs.~\cite{AMM99,Marin,Cuevas}). Thus, they can be expressed in terms of a truncated Fourier series expansion:

\begin{equation}\label{eq:series}
    u_n(t)=\sum_{k=-k_m}^{k_m} z_{k,n}\exp(i k (\wb t+\alpha))
\end{equation}

\noindent with $k_m$ being the maximum of the absolute value of the running index $k$. In the numerics, $k_m$ has been chosen as 13. After the introduction of (\ref{eq:series}), the dynamical equations (\ref{eq:dyn}) transform into a set of $N\times(2k_m+1)$ algebraic equations where the variables are $Z\equiv\{z_{k,n}\}$:

\begin{equation}\label{eq:Fourier}
    F_{k,n} \equiv-\wb^2k^2z_{k,n}+\F_k[V'(u_n)]+\ee(2z_{k,n}-z_{k,n-1}-z_{k,n+1})=0
\end{equation}

\noindent Here, $\F_k$ denotes the Discrete Fourier Transform:

\begin{equation}
    \F_k[V'(u_n)]=\frac{1}{2k_m+1}\sum_{j=-k_m}^{k_m}V'(u(t_j))\exp(-i k \wb t_j),
\end{equation}

\noindent where $t_j$ is a sample of times that must be chosen equally spaced:

\begin{equation}
    t_j=\frac{2\pi j}{\wb(2k_m+1)}, \qquad j=-k_m,\ldots,+k_m,
\end{equation}

\noindent and $u_n(t_j)$ is calculated from the Fourier coefficients $z_{k,n}$ by means of the Inverse Discrete Fourier Transform:

\begin{equation}
    u_n(t_j)=\sum_{k=-k_m}^{k_m}z_{k,n}\exp(i k \wb t_j).
\end{equation}

\begin{figure}
\begin{center}
\begin{tabular}{cc}
    \includegraphics[width=\middlefig]{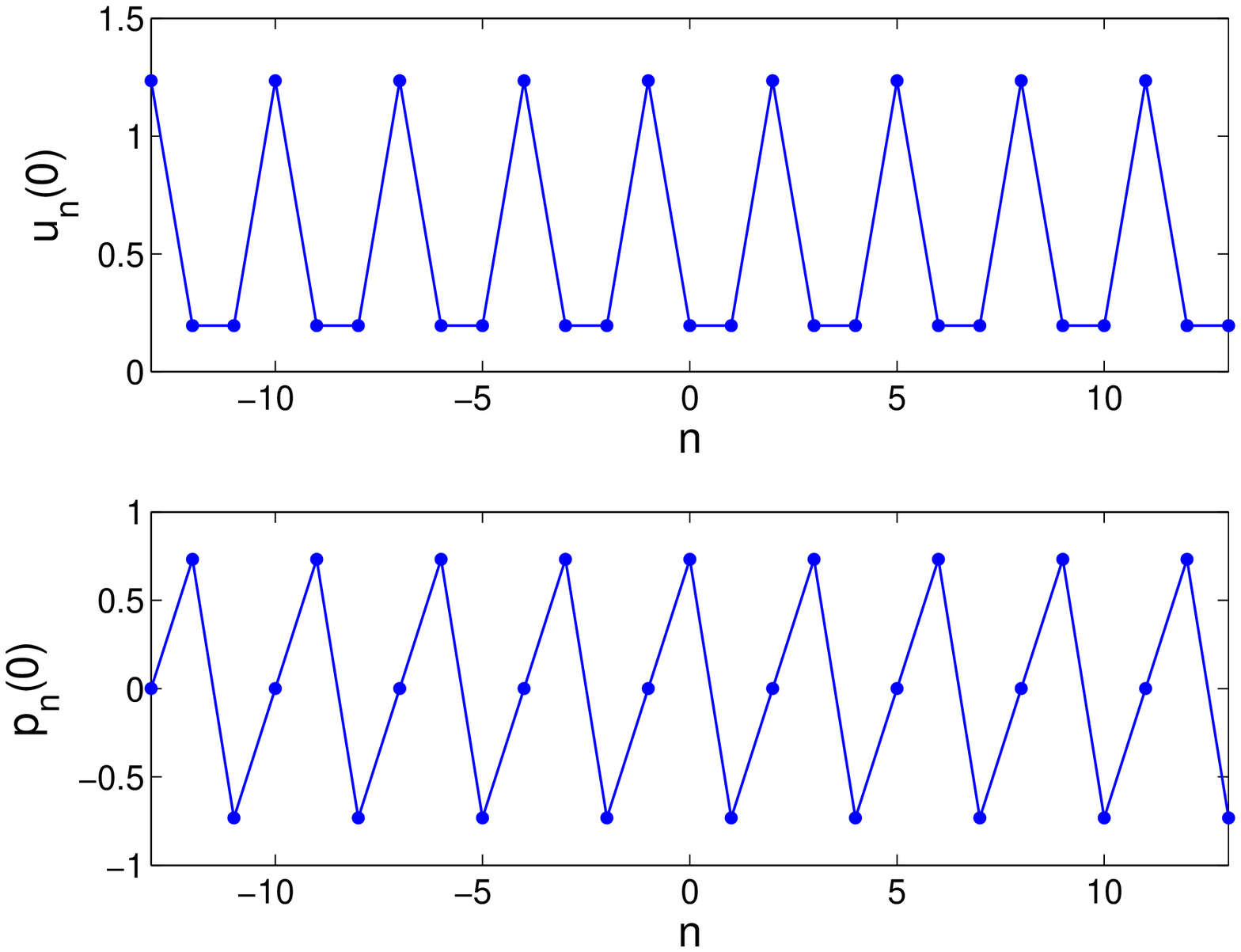} &
    \includegraphics[width=\middlefig]{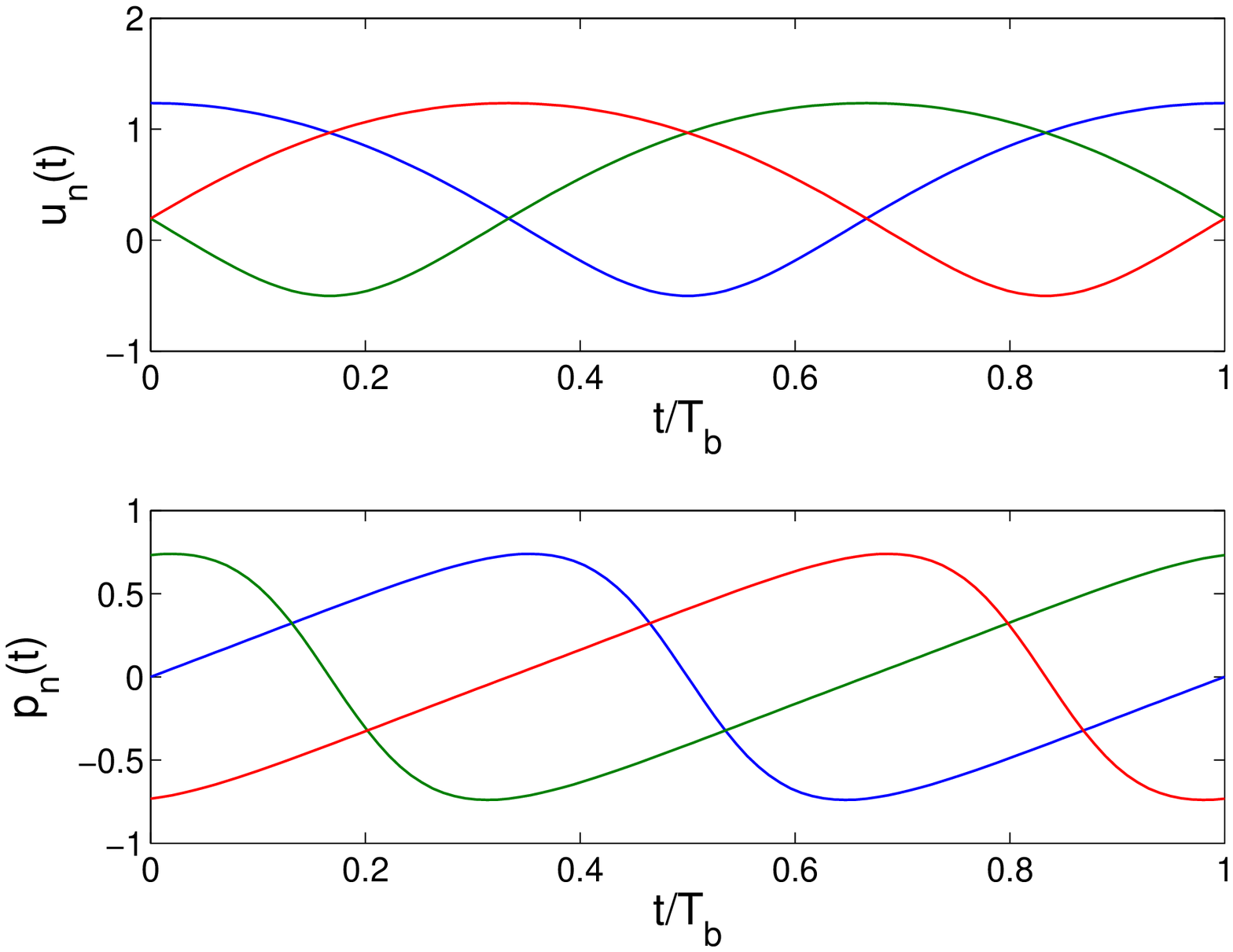} \\
\end{tabular}
\caption{Position $u_n(t)$ and linear momentum $p_n(t)=\dot u_n(t)$ profiles (left panel) and time evolution (right panel) for a phonobreather with $\wb=0.8$, $\ee=0.05$, $\a=2\pi/3$ and a Morse potential. Different curves in right panels stand for different lattice sites. Notice that there are only 3 curves in those panels because this is the spatial periodicity of the phonobreather.} \label{fig:breather1}
\end{center}
\end{figure}

\begin{figure}
\begin{center}
\begin{tabular}{cc}
    \includegraphics[width=\middlefig]{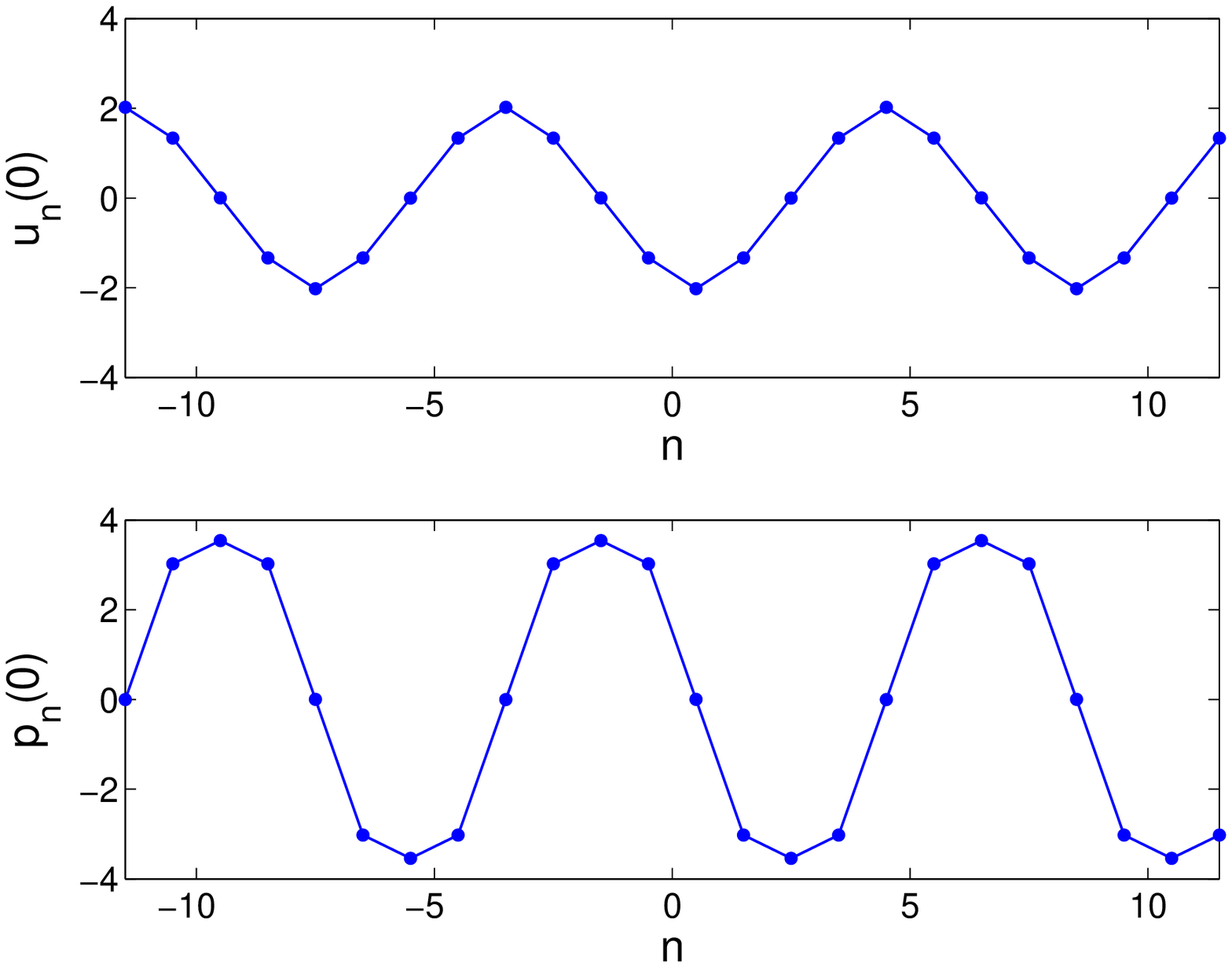} &
    \includegraphics[width=\middlefig]{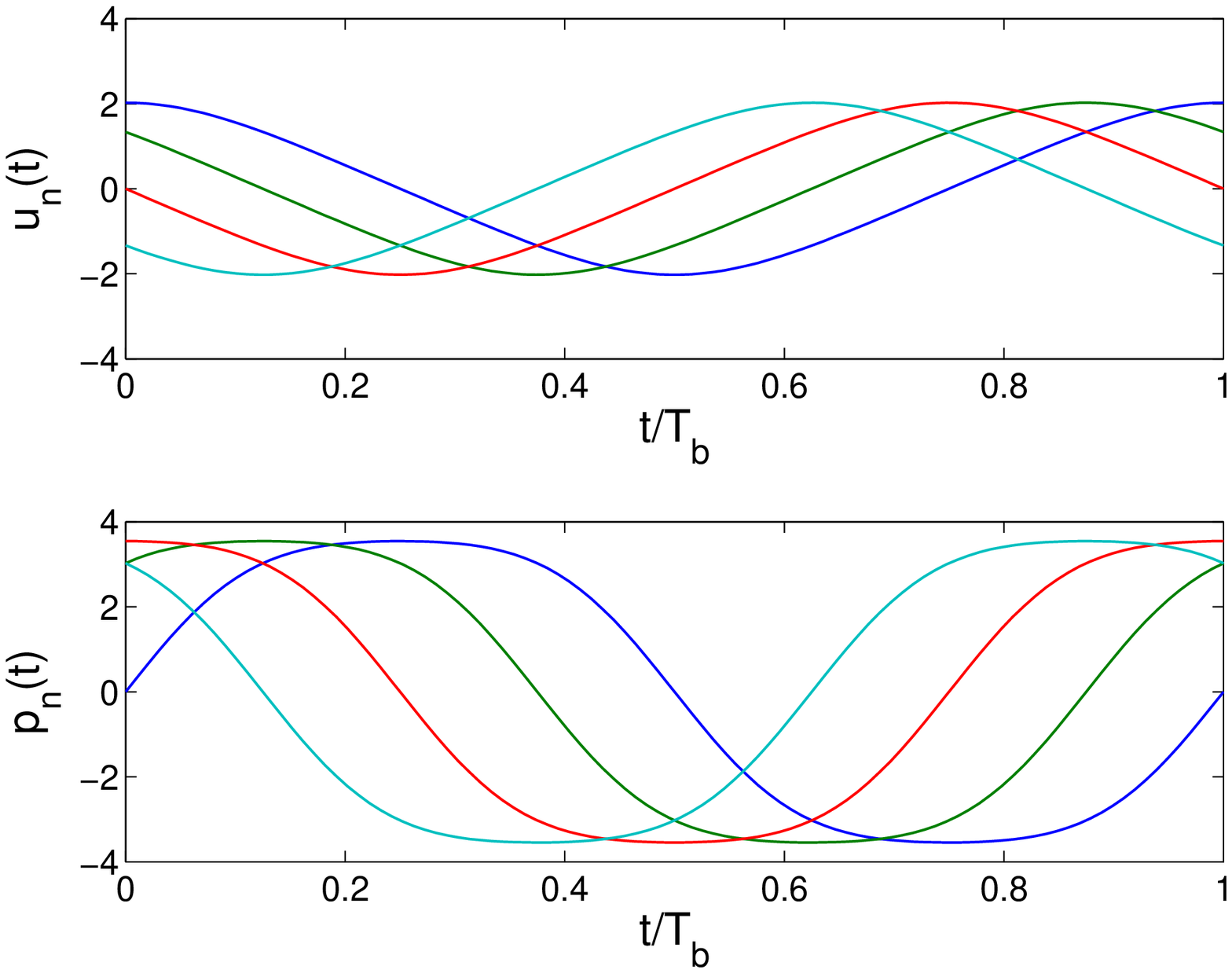} \\
\end{tabular}
\caption{Position $u_n(t)$ and linear momentum $p_n(t)=\dot u_n(t)$ profiles (left panels) and time evolution (right panel) for a phonobreather with $\wb=2$, $\ee=0.05$, $\a=\pi/4$ and a $\phi^4$ potential. Different curves in right panels stand for different lattice sites. Notice that there are only 4 curves in those panels because this is the spatial periodicity of the phonobreather.} \label{fig:breather2}
\end{center}
\end{figure}

Fourier space methods provide with an analytical form of the Jacobian $\J\equiv\partial F/\partial Z$, whose elements are $\{\partial F_{k,n}/\partial z_{k',n'}\}$. Figs. \ref{fig:breather1} and \ref{fig:breather2} show the profiles and time evolution of two examples of phonobreathers \footnote{Notice that hereby we will omit, for simplicity, the non-time reversible character of the solutions} with different phases and potentials.

Phonobreathers can transport energy by means of the phase torsion mechanism, and consequently, there is a stationary energy flux as long as $\alpha\neq0,\pi$  \cite{A97}. With the aid of the Fourier series expansion, the energy flux can be expressed as:

\begin{equation}
    J_{n\rightarrow m}=-J_{m\rightarrow n}=
    \frac{\epsilon}{\Tb}\int_0^{\Tb} u_n(t)\dot{u}_m(t)\dd t=
    -i\wb\epsilon\sum_kkz_{k,n}z_{-k,m}
\end{equation}

Phonobreathers are characterized by a constant energy density at each lattice site. Thus, it must be fulfilled that $z_{k,n+1}=z_{k,n}\exp(ik\alpha)$ with $z_{k,n}=z^*_{-k,n}$ and the flux between neighbouring sites is homogeneous and fulfills:

\begin{equation}\label{eq:flux}
    J_0\equiv J_{n,n+1}=2\wb\ee\sum_{k\geq1} k |z_{k,n}|^2\sin(k\alpha) \qquad\forall n
\end{equation}

\subsection{Linear stability equations}

In order to study the linear stability of phonobreathers, we
introduce a small perturbation $\xi_n$ to a given solution
$u_{n0}$ of Eq. (\ref{eq:dyn}) according to $u_n=u_{n0}+\xi_n$.
Then, the equations satisfied to first order on $\xi_n$ is:

\begin{equation}\label{eq:stab}
    \ddot\xi_n+V''(u_{n0})\xi_n+\ee(2\xi_n-\xi_{n+1}-x_{n-1})=0
    \,,
\end{equation}

\noindent or, in a more compact form:

\begin{equation}\label{eq:Newton0}
    \mathcal{N}(\{u(t)\})\xi=0\,,
\end{equation}

\noindent where $\mathcal{N}(\{u(t)\})$ is known as the Newton
operator. In order to study the orbital stability analysis of the
relevant solution, a Floquet analysis can be performed if there
exists $T_b\in\mathbb{R}$ so that the map $u_n(0)\rightarrow
u_{n}(\Tb)$ has a fixed point \cite{A97}. Then, the stability
properties are given by the spectrum of the Floquet operator
$\mathcal{M}_0$ (whose matrix representation is the monodromy)
defined as:

\begin{equation}\label{eq:monodromy}
    \left(\begin{array}{c} \{\xi_{n}(\Tb)\} \\ \{\dot\xi_{n}(\Tb)\} \\ \end{array}
    \right)=\mathcal{M}_0\left(\begin{array}{c} \{\xi_{n}(0)\} \\ \{\dot\xi_{n}(0)\} \\ \end{array}
    \right)
\end{equation}

The $2N\times2N$ monodromy eigenvalues $\Lambda=\exp(\ii\theta)$
are dubbed the {\em Floquet multipliers}. This operator is
symplectic and real, which implies that there is always a pair of
multipliers (corresponding to the phase and growth modes) at $1$
and that the eigenvalues come in duplets $\{\Lambda,1/\Lambda\}$
if they are real or quadruplets
$\{\Lambda,1/\Lambda,\Lambda^*,1/\Lambda^*\}$ if they are complex.
Consequently, if the phonobreather is stable, all the eigenvalues
lie on the unit circle.

Equation (\ref{eq:Newton0}) can be seen as the eigenequation of
the Newton operator for the eigenvalue $E=0$. Then, the
eigenequations for the Newton operator are:

\begin{equation}\label{eq:bands}
    \mathcal{N}(\{u(t)\})\xi=E\xi \rightarrow \ddot\xi_n+V''(u_{n0})\xi_n+\ee(2\xi_n-\xi_{n+1}-x_{n-1})=E\xi_n
\end{equation}

The Newton operator is periodic in time, and, consequently, its
eigenvectors fulfill the Floquet-Bloch theorem. This theorem
implies that the $E$-eigenvalues spread bringing about a set of
bands $\E_\nu=\E_\nu(\theta)$, where $\theta$ can be chosen in the
First Brioullin zone, i.e. $\theta\in[-\pi,\pi]$. The set of
eigenvalues $E_\nu(\theta)$ with $\theta$ real  is denoted as the
$\nu$-th band. The bands are associated to stable solutions
as long as $\theta_\nu\in\mathbb{R}$. The values of $\theta(E)$
can be obtained by diagonalizing the matrix $\mathcal{M}_E$, which
is obtained in a similar fashion to Eq. (\ref{eq:monodromy}) but
integrating Eq. (\ref{eq:bands}) for each value of $E$. The
monodromy corresponds obviously to $E=0$ and, consequently, the
Floquet arguments correspond to $\theta_\nu(0)$. Thus,  a solution
is stable if there are be $2N$ bands that either cross the $E=0$
axis or are tangent to it. More details on this theory, called
{\em Aubry's band theory} can be found in Ref.~\cite{A97}.

\section{Analytical results}\label{sec:MST}

In this section, we will show some analytical predictions about
the stability of phonobreathers at low coupling. To this end, we
start by recalling previous results established by some of
the authors for multibreathers with an arbitrary number of excited
sites and continue by applying these results to the prediction of
the stability properties of phonobreathers with Morse and $\phi^4$
potentials at low coupling.

\subsection{Previous results}

For the sake of completeness, we recall in this subsection some
previous results on the Multibreathers Stability Theorem (MST)
proposed in Ref.~\cite{ACSA03}. For more details, the reader is
also referred to Refs.~\cite{CAR05,Arxiv}.

The MST refers to Klein-Gordon systems of the form (\ref{eq:dyn})
and estimates the displacement experienced by Aubry's bands  when
the coupling parameter $\ee$ is switched on.

Suppose that $u_n^0$ is a $\Tb$-periodic solution at the
anti--continuous limit ($\ee=0$), with $p$ excited oscillators and
$N-p$ ones at rest ($u_n^0=0$). At this limit, there are $p$
degenerated bands tangent to the $E=0$ axis at $(E,\theta)=(0,0)$.
Their curvature is positive (negative) for soft (hard) on-site
potentials. The MST can predict the displacement of the minimum of
this bands $\{\Delta E\}_{i=0}^{p-1}=\{E\}_{i=0}^{p-1}=\ee\{\la\}_{i=0}^{p-1}$, with
$\{\la\}$ being the set of eigenvalues  for the perturbation
matrix $Q$. The (non-zero) non-diagonal elements of the $p\times
p$ perturbation matrix $Q$ in reduced form (see below) are defined
as

\begin{equation}
    Q_{n,n\pm1}\,=\,-\,\frac{1}{\mu_n\, \mu_{n\pm1}}\int_{-\Tb/2}^{\Tb/2}\,\dot{u}^0_n
    \dot{u}_{n\pm1}^0\,\dd t \label{eq:qnmD}
\end{equation}
with $\mu_n=\sqrt{\int_{-T/2}^{T/2}(\dot{u}_n^0)^2\dd t}$. Only
the indexes corresponding to the excited oscillators are
considered and they are renumbered them from $1$ to $p$. If we
considered all the oscillators, the matrix $Q$ would be in full
form. Each oscillator at rest adds a row and a column of zeros
and, therefore, a zero eigenvalue, which is not relevant for the
stability properties.

The diagonal elements are given by

\begin{equation}
Q_{n,n}=-\frac{\mu_{n+1}Q_{n,n+1}+\mu_{n-1}Q_{n,n-1}}{\mu_n} \,.
\label{eq:qnnD}
\end{equation}

With these definitions, we reproduce here the multibreather stability theorem:
\\ \mbox{}\\
\noindent {\bf Generalized MST} {\em Given a Klein--Gordon system, Eq.~(\ref{eq:dyn}), a specific multibreather
solution at zero coupling $\{u^0_n\}$, $u(t)$ the corresponding
solution at low and positive coupling, $\{\lambda\}_{i=0}^{p-1}$
the eigenvalues of the reduced, perturbation matrix $Q$, with only
one zero, then:
\\The solution $u(t)$ is stable if:
\\a)~The on--site potentials are soft and there is not any positive value in $\{\lambda_i\}_{i=0}^{p-1}$.
\\b)~The on--site potentials are hard and there is not any negative value in $\{\lambda_i\}_{i=0}^{p-1}$.
 }
\\ \mbox{}\\

We can summarize the stability properties in the following way. If $S=1$ means stability and $S=-1$ instability, $H=1$ corresponds to a hard on-site potential, and $H=-1$ to a soft one, and we define $\sign(Q)=1$ if all the eigenvalues of $Q$ but a zero one are positive and $\sign(Q)=-1$ if they are negative except for the zero one, then

\begin{equation}
    S=H\times \sign(Q). \label{eq:stability}
\end{equation}
It there are eigenvalues of different signs, the multibreather is
always unstable. It is important to take into account that there
is always a zero eigenvalue due to a global phase mode. If there
is more than one zero eigenvalue, the stability theorem can only
predict the instability in the case that there exists at least one
eigenvalue $\lambda$ that leads to $S=-1$ in the previous equation
(changing $\sign(Q)$ for $\sign(\lambda)$),  but not the
stability, as the 0--eigenvalue is degenerate.


\subsection{Application to phonobreathers}

We apply hereby the theory recalled in the previous subsection for the system given by Eq. (\ref{eq:dyn}). Consequently, we must construct the reduced perturbation matrix $Q$ whose elements are given by Eqs. (\ref{eq:qnmD}) and (\ref{eq:qnnD}). The functions $u_n^0(t)$ that appear in those equations are the solutions of the isolated oscillators submitted to the potentials $V(u_n)$, i.e. the solutions of the equations:

\begin{equation}
    \ddot u^0_n+V'(u^0_n)=0,
    \label{eq:isolated}
\end{equation}

Let $I_{nm}$ be defined as:

\begin{equation}\label{eq:Inm}
    I_{n,m}=\int_{-\Tb/2}^{\Tb/2}\,\dot{u}_n^0(t)\,\dot{u}^0_m(t)\,\dd t\,,
\end{equation}

and the parameters $\t_{n,m}$ as:

\begin{equation}\label{eq:thetaint}
    \t_{n,m}=\fracc{I_{n,m}}{\sqrt{I_{n,n}I_{m,m}}}.
\end{equation}

Substituting Eq. (\ref{eq:series}) into (\ref{eq:Inm})-(\ref{eq:thetaint}) and taking into account that the phase difference between neighbouring sites $\alpha$ is constant, we get that

\begin{equation}\label{eq:theta}
    \t_{n,n\pm1}\equiv\t=\frac{\displaystyle \sum_k k^2 z_k^2 \cos(k\a)}{\displaystyle\sum_k k^2z_k^2}.
\end{equation}

With the aid of the $\t$ parameter, the $Q$ matrix can be written as

\begin{equation}
    Q=\t Q^{(R)},
\end{equation}

where $Q^{(R)}$ is the $Q$ matrix for a time-reversible phonobreather (i.e. with $\alpha=0$) in an homogeneous lattice:

\begin{equation}
    Q^{(R)}_{n,m}=2\delta_{n,m}-(\delta_{m,n+1}+\delta_{n,n-1})\,,\qquad n=1,\ldots N
\end{equation}

Thus, the eigenvalues of our problem are given by $\la=\t\la_R$ with $\la_R$ the eigenvalues of the time-reversible lattice. These eigenvalues are given by $\la_{R,n}=4\sin^2(n\pi/N)$ with $n=0,\ldots,N-1$ \cite{ACSA03}; consequently:

\begin{equation}
    \la_n=4\t\sin^2\frac{n\pi}{N} \qquad n=0,\ldots,N-1.
\end{equation}

Thus, $\sgn(\la_n)=\sgn(\t)$. Following Eq. (\ref{eq:stability}),
the phonobreather is stable for a soft potential if $\t<0$ and,
for a hard potential, if $\t>0$. In the case of a symmetric
on-site potential, $\displaystyle z_{2,\nu}=0,
\forall\nu\in\mathbb{Z}$, and consequently, $\t(\pi/2)=0$.

We calculate below the expression of $\t$ for a Morse and a hard $\phi^4$ potential:

\begin{figure}[t]
\begin{center}
\begin{tabular}{cc}
    \includegraphics[width=\middlefig]{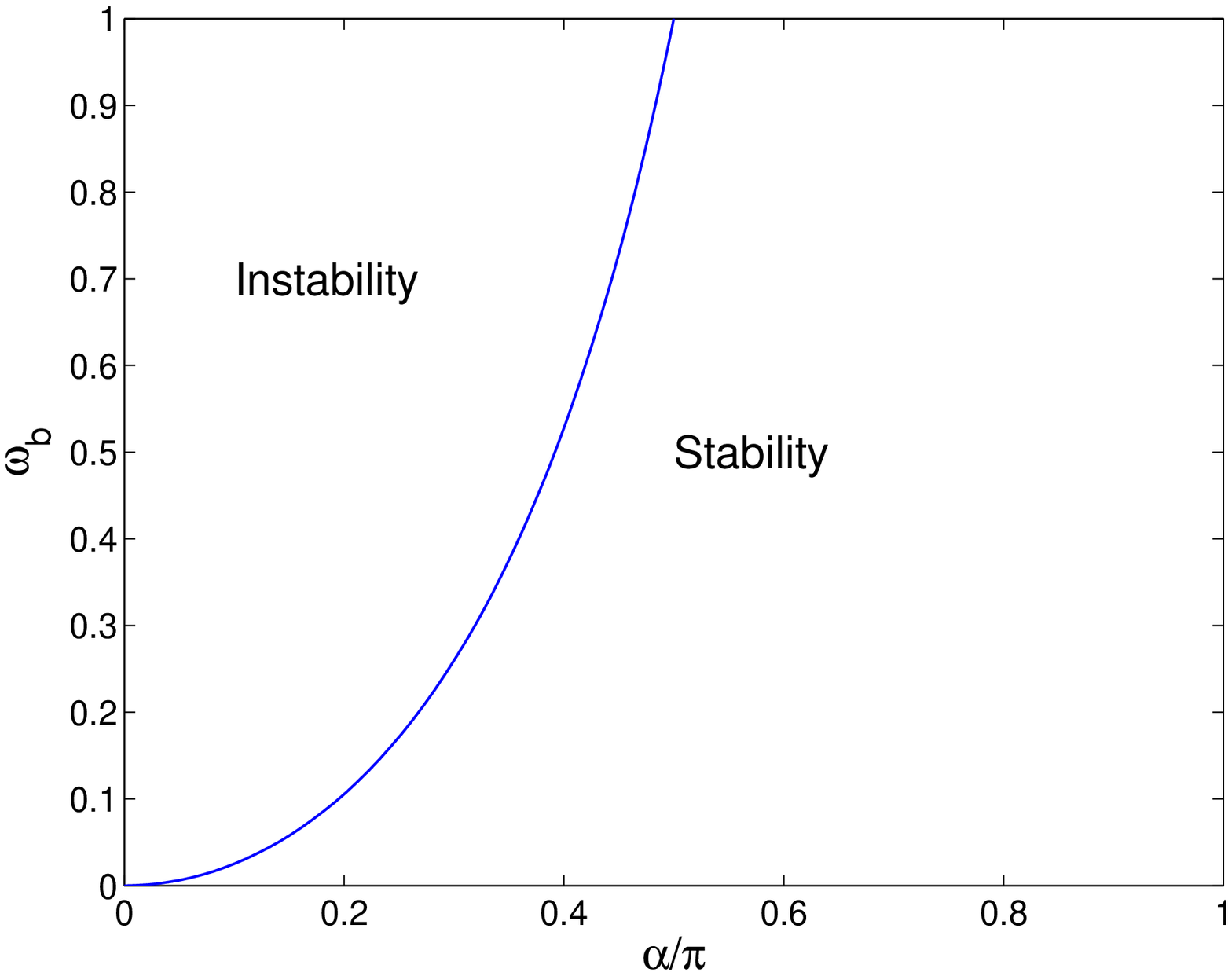} &
    \includegraphics[width=\middlefig]{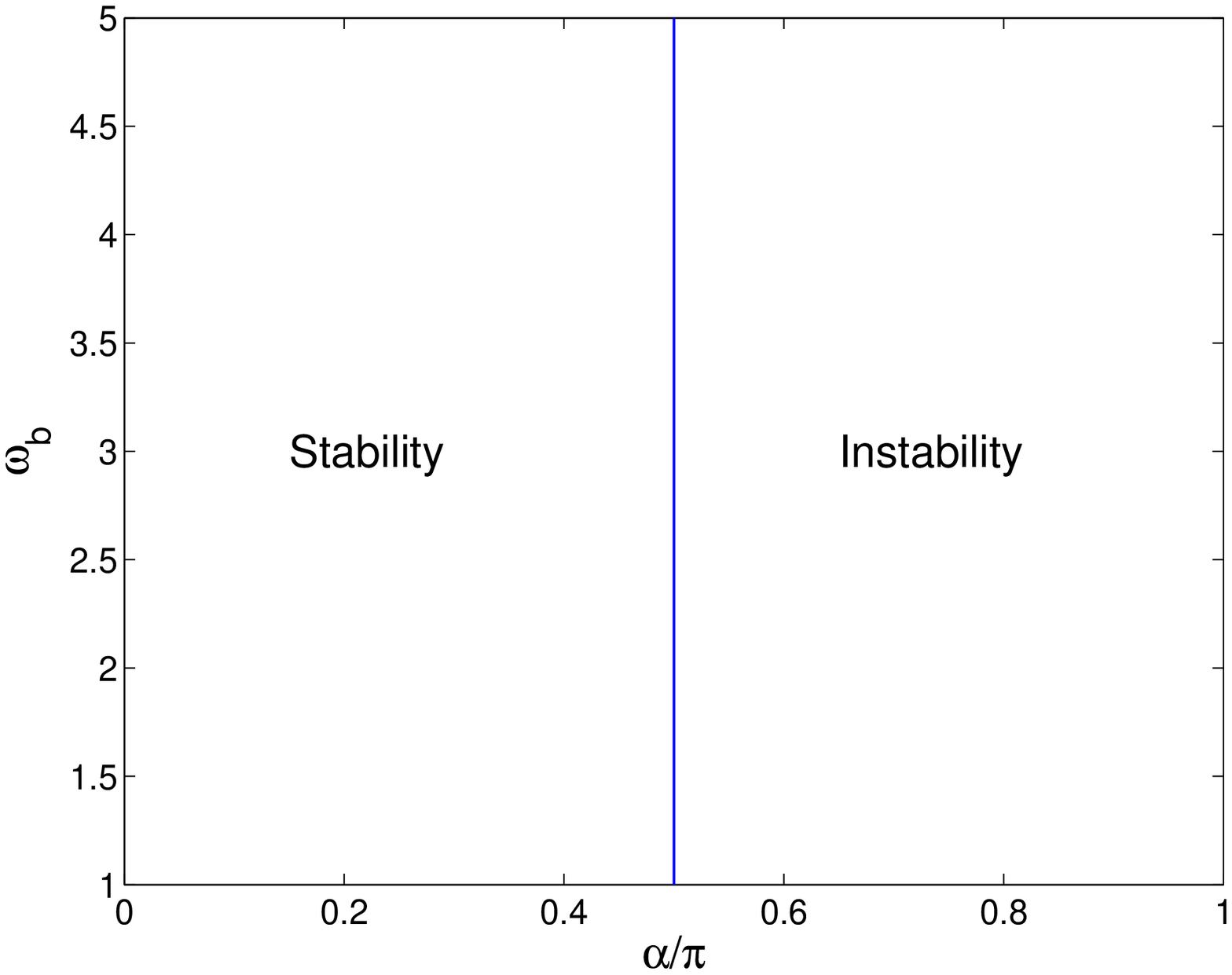} \\
\end{tabular}
\caption{Stability and instability regions predicted by the MST for Morse (left) and $\phi^4$ (right) potentials. Recall that, for the Morse ($\phi^4$) potential, instability implies $\t<0$ ($\t>0$).} \label{fig:theta}
\end{center}
\end{figure}

\subsubsection{Morse potential}

The orbits of Eq. (\ref{eq:isolated}) are given by \cite{ACSA03}:
\begin{equation}\label{xt}
    u(t)=\log\frac{1-\sqrt{1-\wb^2}\cos\wt}{\wb^2}\,,
\end{equation}
The Fourier coefficients are:
\begin{equation}
    z_{0,n}=\log\frac{1+\wb}{2\wb^2} \quad;\quad
    z_{k,n}=-\,\frac{(-1)^{k+1}}{k}\left(\frac{1-\wb}
    {1+\wb}\right)^{k/2}\,,
\end{equation}
Thus, we can write:
\begin{equation}
    \t=\frac{\sum_{k=1}^{\infty}r^k\cos(k\a)}{\sum_{k=1}^{\infty}r^k},
\end{equation}
with
\begin{equation}
    r=\frac{1-\wb}{1+\wb}.
\end{equation}

Using \cite{GR65}, the sums can be performed, leading to:

\begin{equation}\label{eq:thetaM}
    \t=\frac{(1-r)(\cos\a-r)}{1-2r\cos\a+r^2}.
\end{equation}

Thus,

\begin{equation}
    \sgn(\t)=\sgn(\cos\a-r),
\end{equation}

In consequence, $\t<0$ (i.e. the phonobreather is stable) if

\begin{equation}\label{eq:cond1}
    \cos\a<r\left(=\frac{1-\wb}{1+\wb}\right).
\end{equation}

The main consequence of this condition is that, for a given value
of $\a$, there exists a critical value of $\wb$ below which the
phonobreather is {\em unstable}. This critical value is:

\begin{equation}\label{eq:wc}
    \wc=\frac{1-\cos\a}{1+\cos\a}=\tan^2\frac{\a}{2}
\end{equation}

Consequently, as $\wb<1$ for oscillators with Morse potential, if
$\a>\pi/2$ phonobreathers are always stable, independently of the
frequency.

Fig. \ref{fig:theta}(left) shows the predictions of the MST for the
Morse potential as a function of $\wb$ and $\a$. For
$\wb=\wc(\a)$, $\t=0$ and the degeneracy of $\la_n$ cannot be
removed; consequently, the MST would not be able to predict the (in)stability
in that case.

\subsubsection{$\phi^4$ potential}

The orbit of an oscillator submitted to a $\phi^4$ hard potential is given by \cite{KA99}:
\begin{equation}
    u_n(t)=\sqrt{\frac{2\k^2}{1-2\k^2}}\cn\left(\frac{\wb t}{\sqrt{1-2\k^2}},\k\right)=
    \sqrt{\frac{2\k^2}{1-2\k^2}}\cn\left(\frac{2K(\k)}{\pi}\wb t,\k\right),
\end{equation}
where $\cn$ is a Jacobi elliptic function of modulus $\k_n$ and $K(\k)$ is the complete elliptic integral of the first kind defined as:
\begin{equation}
    K(\k)=\int_{0}^{\pi/2}\,\frac{\dd x}{\sqrt{1-\k^2\sin^2x}}\,.
\end{equation}
The breather frequency $\wb$ is related to the modulus $\k$ through:
\begin{equation}
    \wb=\frac{\pi}{2\sqrt{1-2\k^2}K(\k)}.
\end{equation}
In order to calculate $\t$, it is better to use Eq. (\ref{eq:thetaint}). Taking into account that $\t$ depends only on the phase difference, we can write:

\begin{equation}
    J_{nm}=\frac{4\k^2K(\k)\wb^3}{\pi(1-2\k^2)}
    \int_{-2K(\k)}^{2K(\k)} \dd x\ \sn(x)\sn(x+a)\dn(x)\dn(x+a),
\end{equation}
with $a=2K(\k)\a/\pi$. This integral can be evaluated applying
\cite[identity 171]{Elliptic}:

\begin{eqnarray}
    \k^2\sn(x)\dn(x)\sn(x+a)\dn(x+a)=&& \nonumber \\
    -\cs(a)\ns(a)(1+\dn^2(a))+ \cs(a)\ns(a)(\dn^2(x)+\dn^2(x+a))-
    \nonumber \\
    \ds(a)(\cs^2(a)+\ns^2(a))[\mathrm{Z}(x+a)-\mathrm{Z}(x)-\mathrm{Z}(a)],
\end{eqnarray}
where $\mathrm{Z}(x)$ is the Jacobian elliptic Z function
\cite{Lawden}. Then, the value of $\t$ is given by:

\begin{equation}\label{eq:theta4}
    \t=3\frac{\cs(a)\ns(a)[2E(\k)-K(\k)(1+\dn^2(a))]-\ds(a)(\cs^2(a)+\ns^2(a))\mathrm{Z}(a)}
    {(-1+2\k^2)E(\k)+(1-\k^2)K(\k)}
\end{equation}

\noindent where $E(\k)$ is the complete elliptic function of the
second kind:
\begin{equation}
    E(\k)=\int_{0}^{\pi/2}\,\sqrt{1-\k^2\sin^2x}\ \dd x\,.
\end{equation}

Three relevant values of $\t$ correspond to $\a=\pi$, $\a=\pi/2$, $\a=0$ which lead to $\t=1$, $\t=0$ and $\t=-1$, respectively. This result can be straightforwardly determined from Eq. (\ref{eq:theta}). An analytic determination of the sign of $\t$ for other values of $\alpha$ is not possible. However, it is possible to visualize numerically that the only zero of $\t(\a)$ takes place at $\a=\pi/2$ being $\t<0$ $(\t>0)$ for $\a<\pi/2$ $(\a>\pi/2)$. In consequence, phonobreathers with a $\phi^4$ hard potential are stable as long as $\a<\pi/2$.

Fig. \ref{fig:theta}(right) summarizes the predictions of the MST for the $\phi^4$ potential as a function of $\wb$ and $\a$. For $\a=\pi/2$, $\t=0$ and the degeneracy of $\la_n$ cannot be removed, and consequently, the MST could not predict (in)stability.

\section{Numerical results}\label{sec:stab}

The analytical results of the previous section concern to an infinitesimally small coupling. In this section, we explore the existence and stability properties for an arbitrary coupling constant and determine the range of validity of the predictions of the multibreather stability theorem.

\subsection{Morse potential}

\begin{figure}
\begin{center}
\begin{tabular}{cc}
    \includegraphics[width=\middlefig]{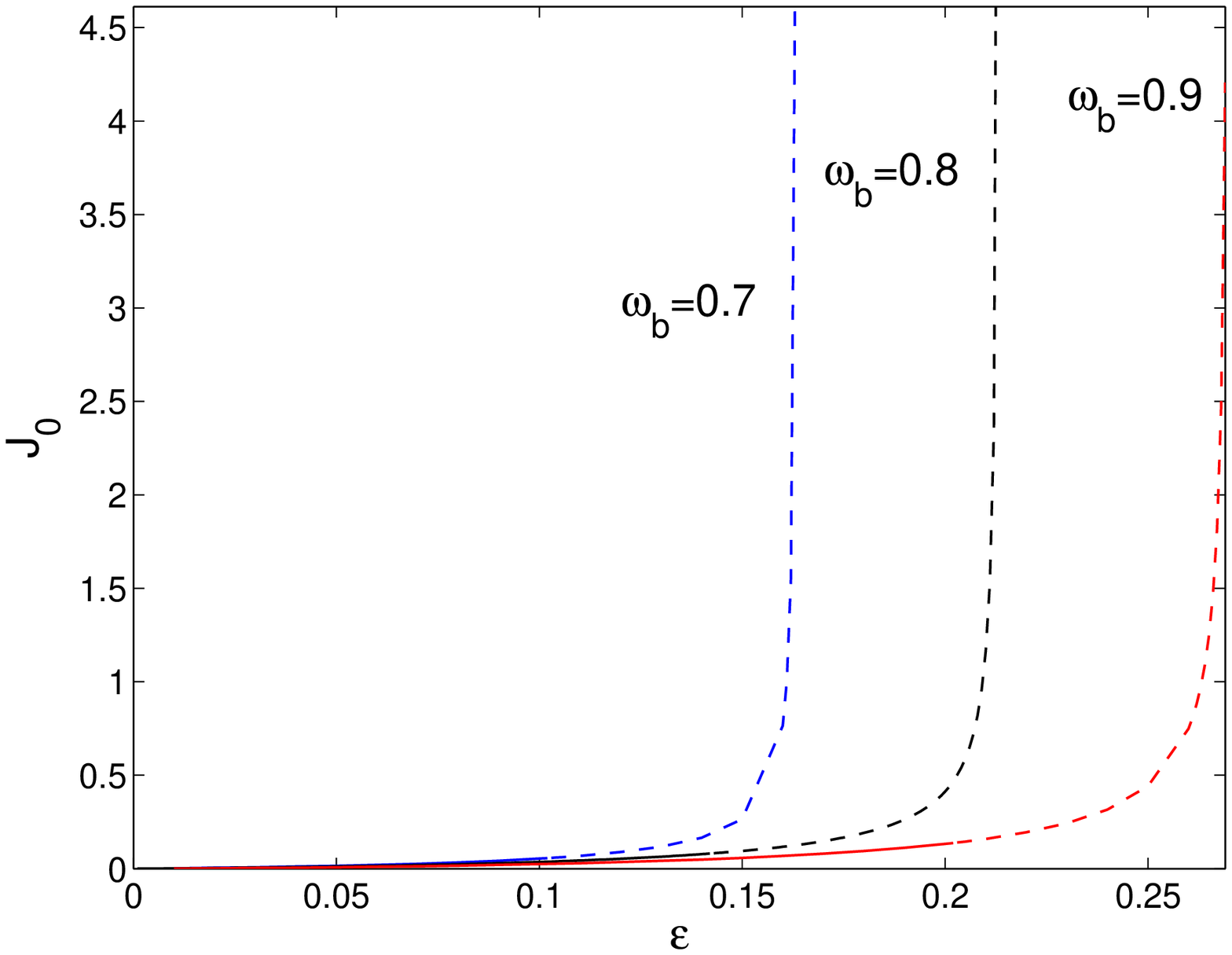} &
    \includegraphics[width=\middlefig]{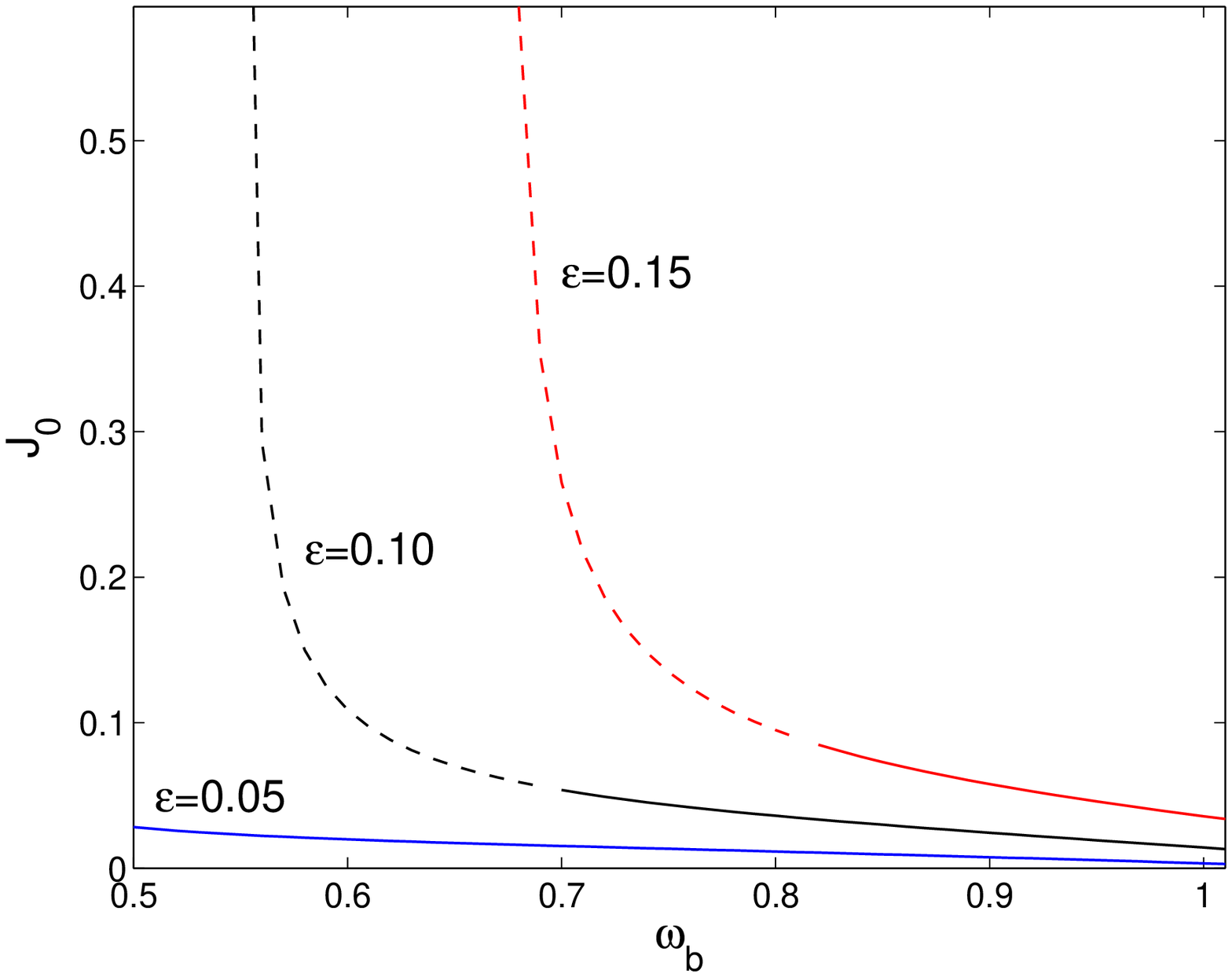} \\
\end{tabular}
\caption{Flux between nearest-neighbours for phonobreathers with $\a=2\pi/3$ and a Morse potential. Dashed lines represent unstable solutions, for which the flux diverges above (below) a critical value of $\ee$ ($\wb$).} \label{fig:flux1}
\end{center}
\end{figure}

Before undertaking the existence and stability analysis of the
phonobreather, we show the properties of the energy flux. Fig.~\ref{fig:flux1} depicts the dependence of $J_0$ with respect to
$\epsilon$ and $\wb$ for $\a=2\pi/3$; the qualitative behaviour
displayed in the figure is generic for any value of $\a$. It is
observed that for the Morse potential, the flux increases
(decreases) with the coupling constant (frequency). There is a
critical value of $\ee$ ($\wb$) above (below) which phonobreathers
do not exist. For this critical value, the flux diverges. This
fact is used in \ref{app:crit} to calculate the
dependence of the critical value of $\ee$ with $\wb$.

Additionally, a full stability analysis has been performed by
varying $\ee$ and $\wb$ for a Morse potential and choosing two
different values of the phase difference $\alpha=2\pi/3$ and
$\alpha=4\pi/9$. In the first case, phonobreathers are stable for
small $\ee$ and become unstable above a critical value of $\ee$.
In the second case, phonobreathers are unstable for small $\ee$ as
long as $\wb>\wc$; the MST predicts that $\wc=0.7041$, which fits
quite well with the numerics; in addition, for large enough $\ee$,
the phonobreather becomes stable, losing the stability again when
$\ee$ is  increased further. As explained above, phonobreathers
cease to exist above a critical value of $\ee$ because the flux
diverges. This critical value of $\ee$ has been calculated in Eq.
(\ref{eq:critMorse}):

\begin{equation}
    \ee_c=\frac{\wb^2}{4\sin^2(\alpha/2)}
\end{equation}

\begin{figure}
\begin{center}
\begin{tabular}{cc}
    \includegraphics[width=\middlefig]{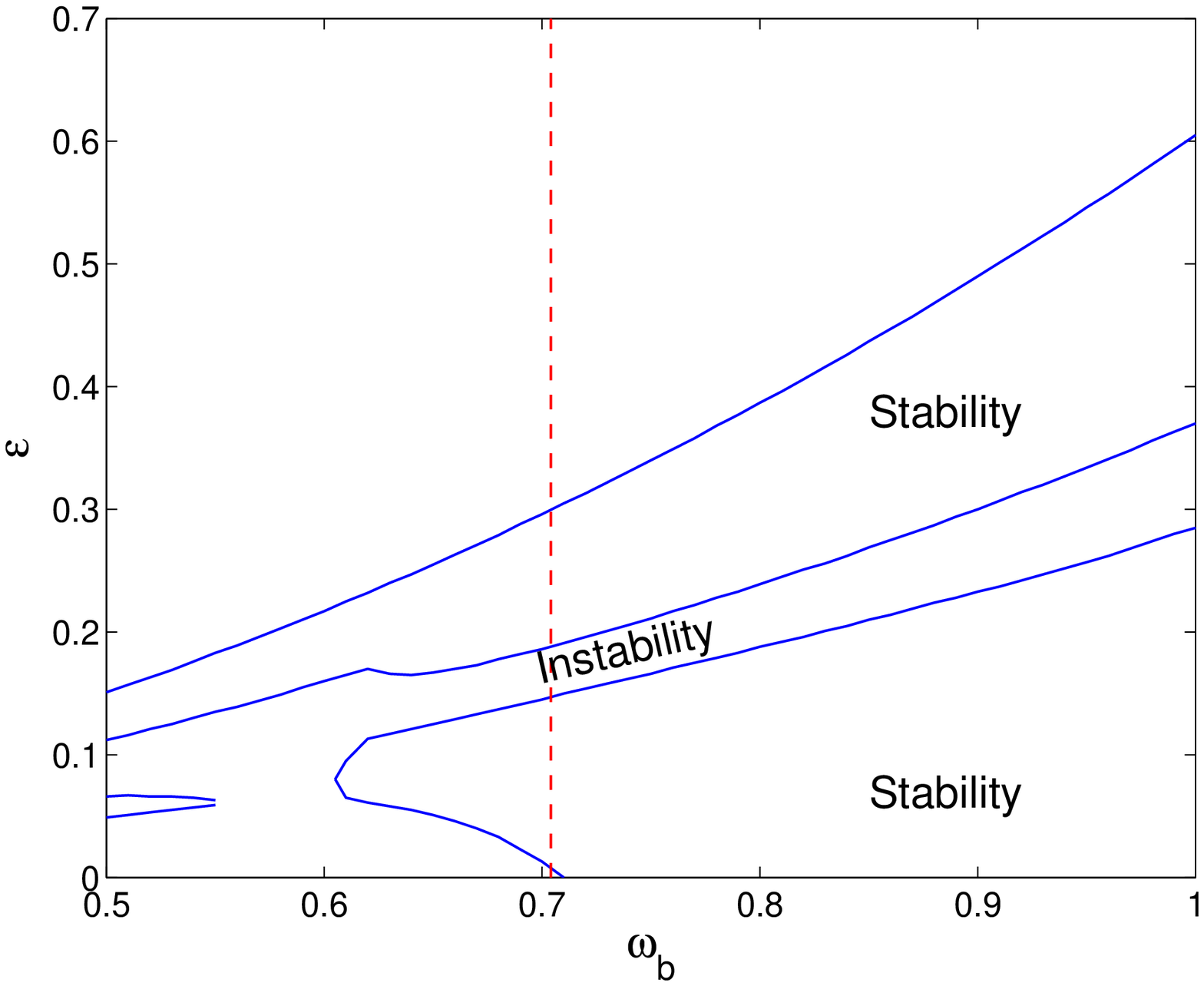} &
    \includegraphics[width=\middlefig]{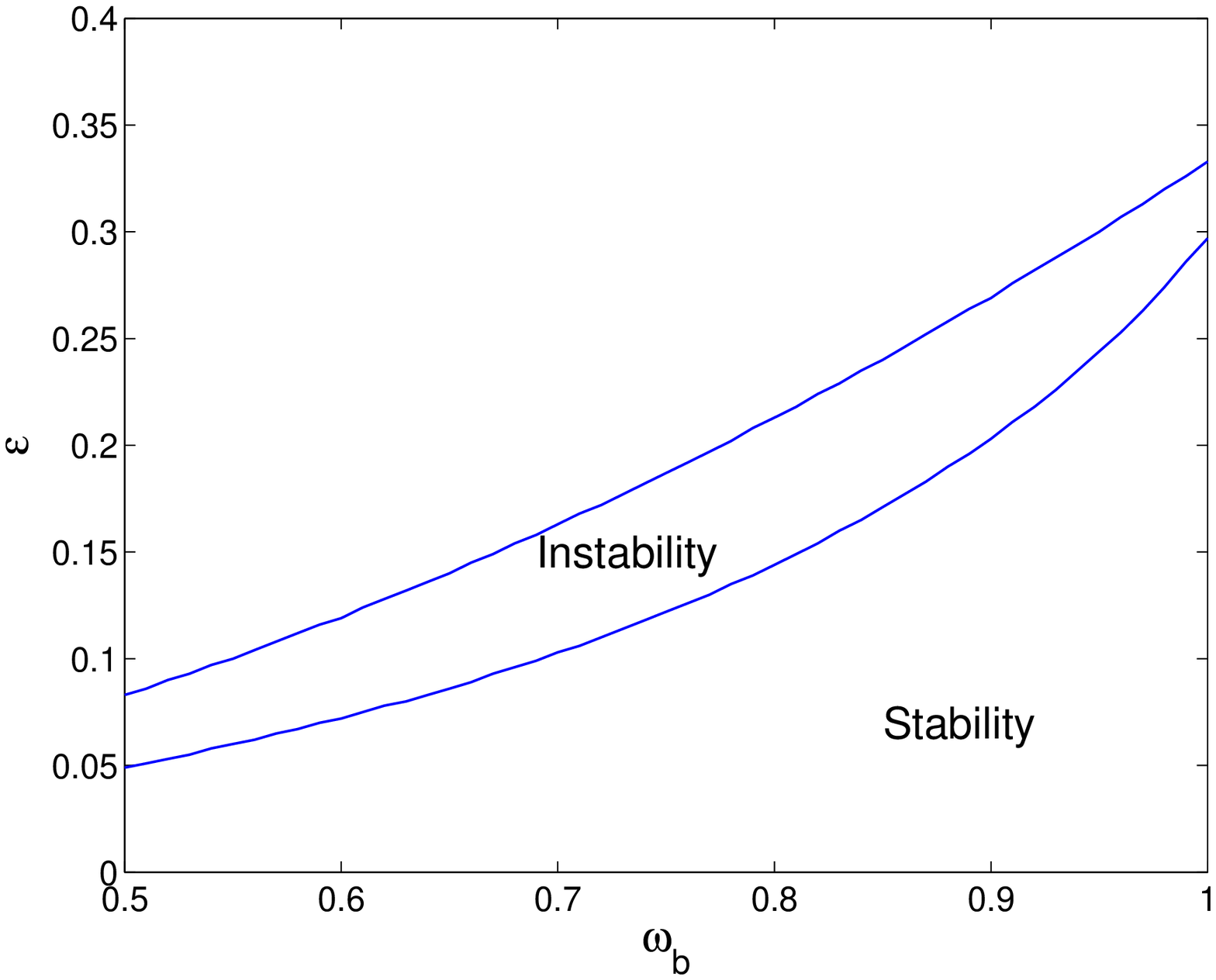} \\
\end{tabular}
\caption{($\ee$-$\wb$) plane showing the existence and stability of
phonobreathers with $\alpha=4\pi/9$ (left) and $\alpha=2\pi/3$ (right) in a Morse potential. Dashed line in left panel corresponds to the MST prediction (i.e. stable at the right of the line and unstable at the left) for low coupling.} \label{fig:planes}
\end{center}
\end{figure}

\begin{figure}
\begin{center}
\begin{tabular}{cc}
    \includegraphics[width=\middlefig]{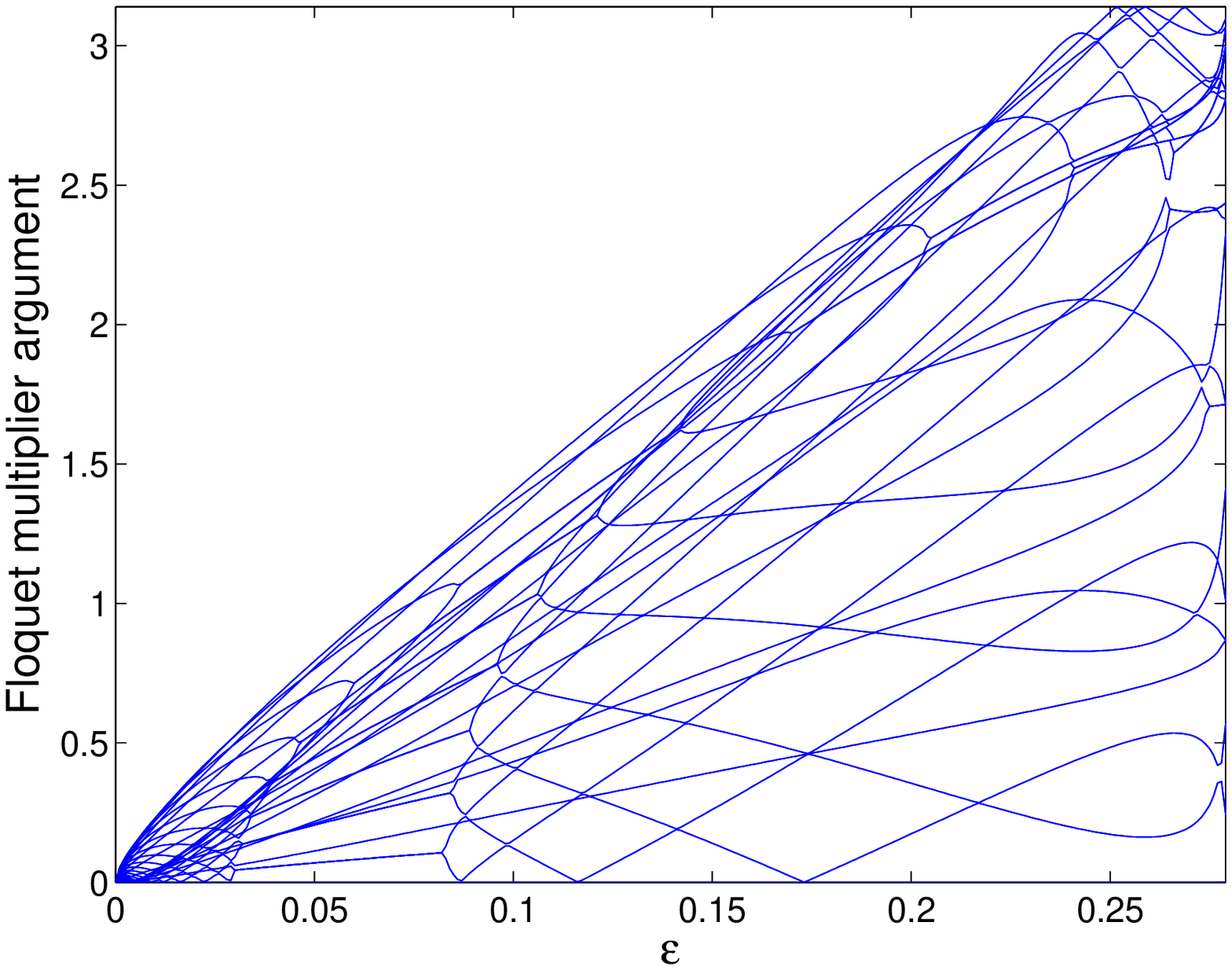} &
    \includegraphics[width=\middlefig]{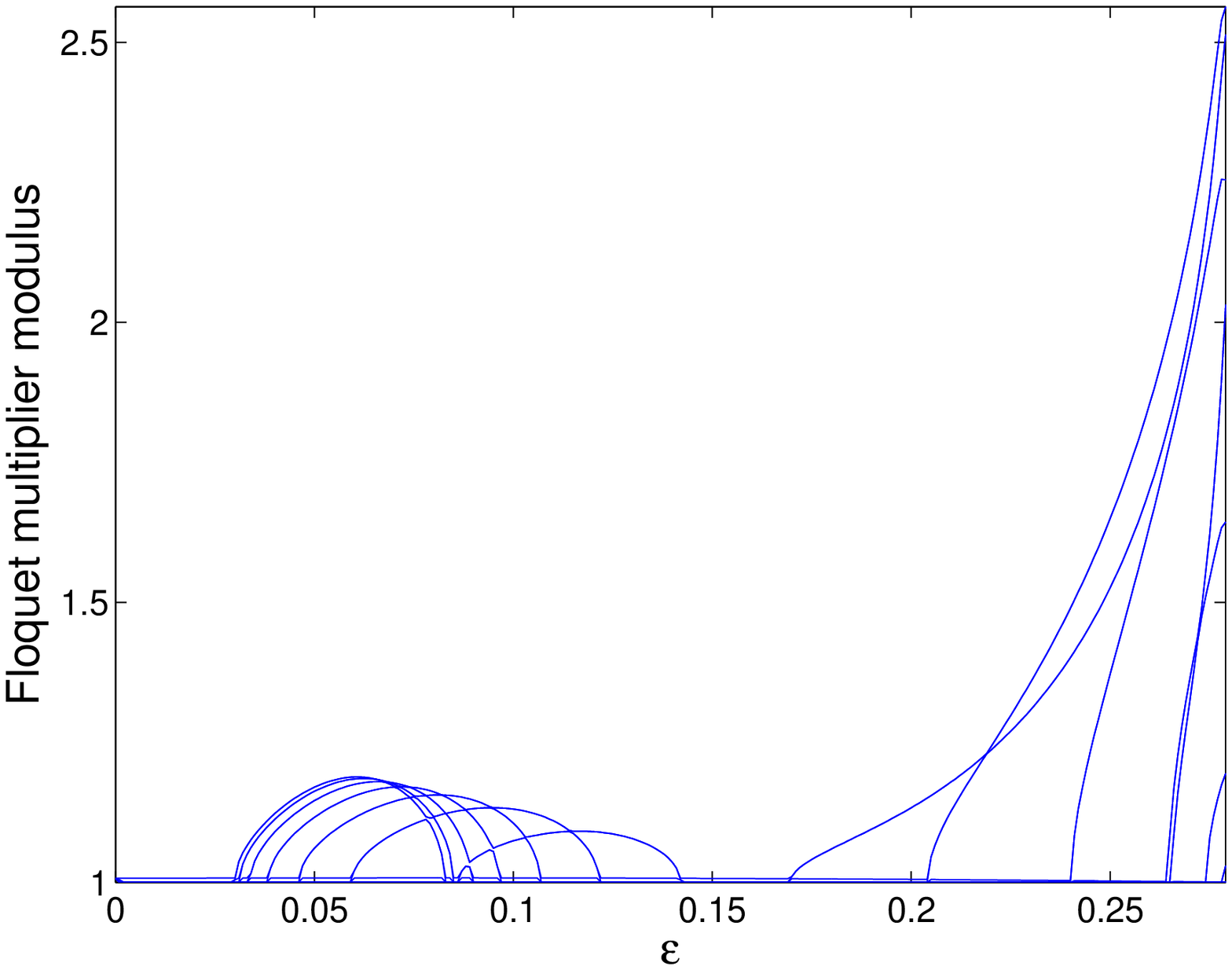} \\
    \includegraphics[width=\middlefig]{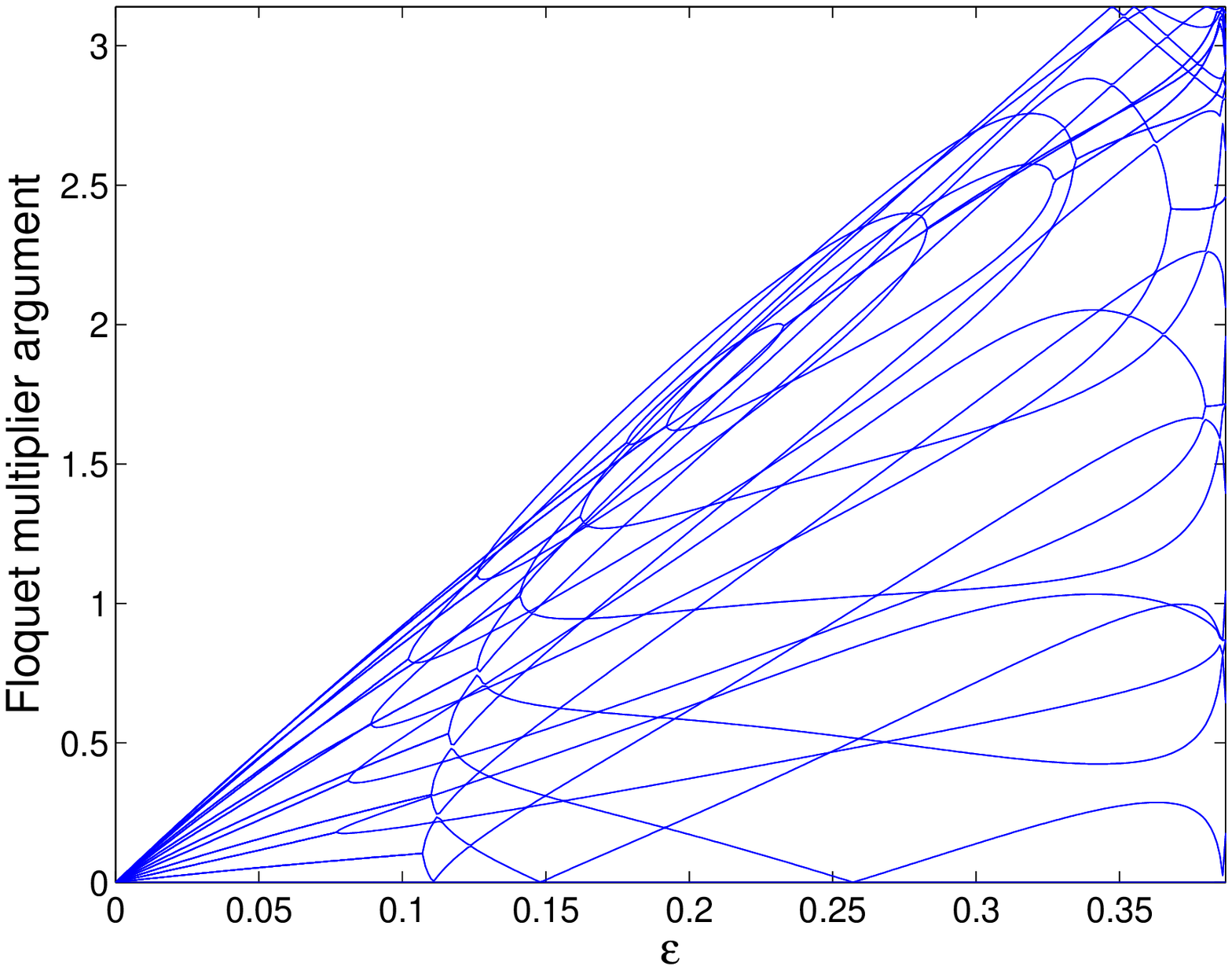} &
    \includegraphics[width=\middlefig]{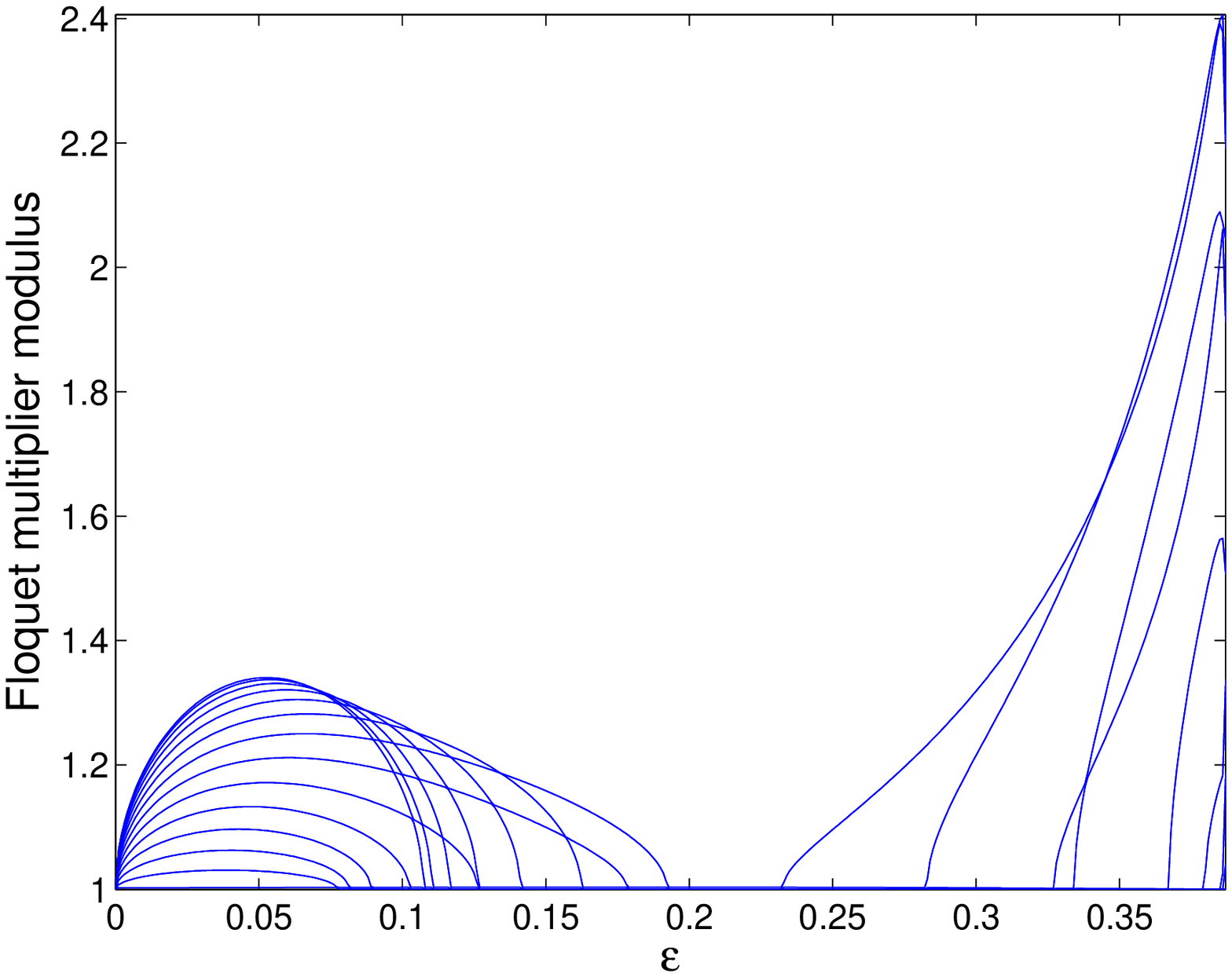} \\
    \includegraphics[width=\middlefig]{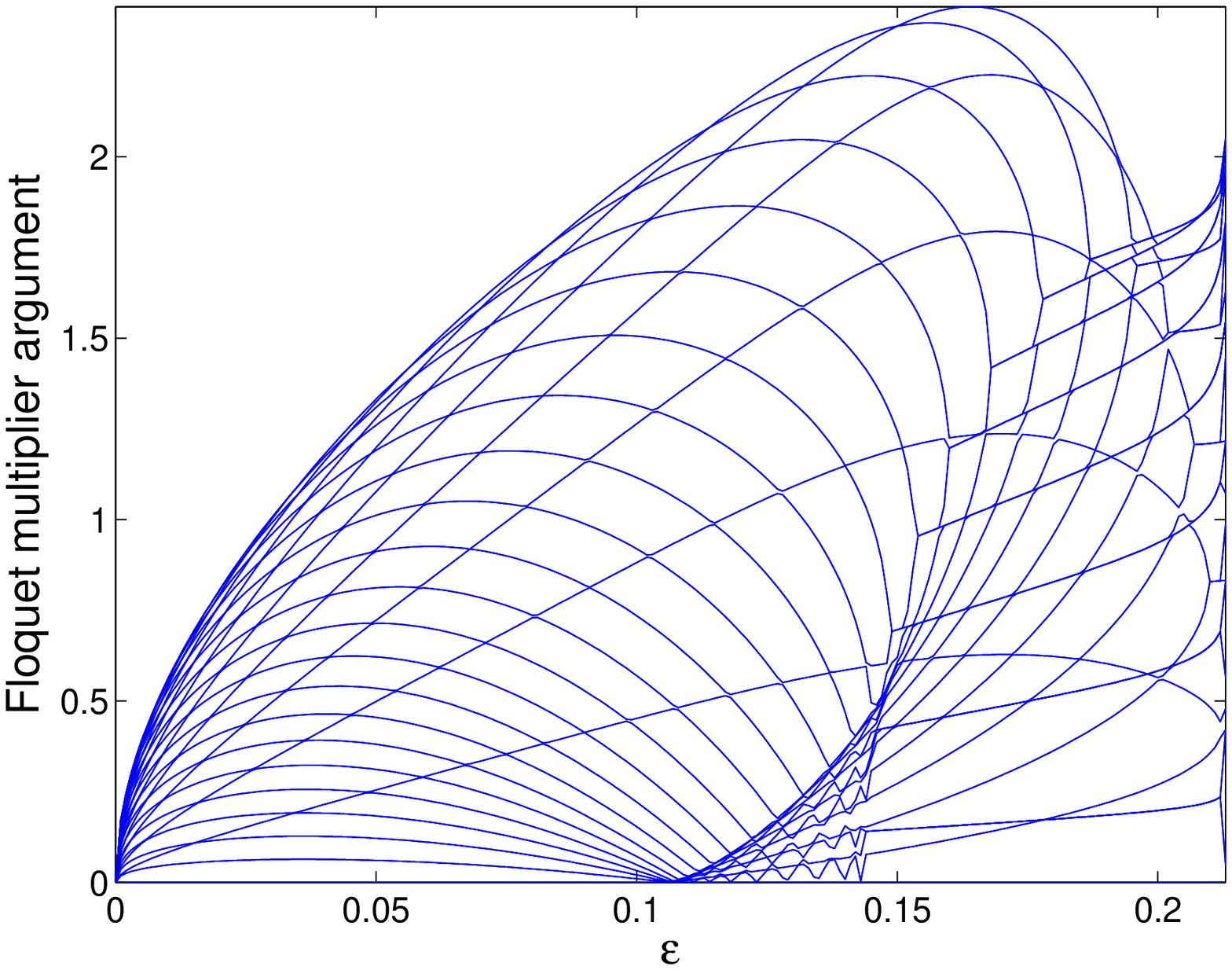} &
    \includegraphics[width=\middlefig]{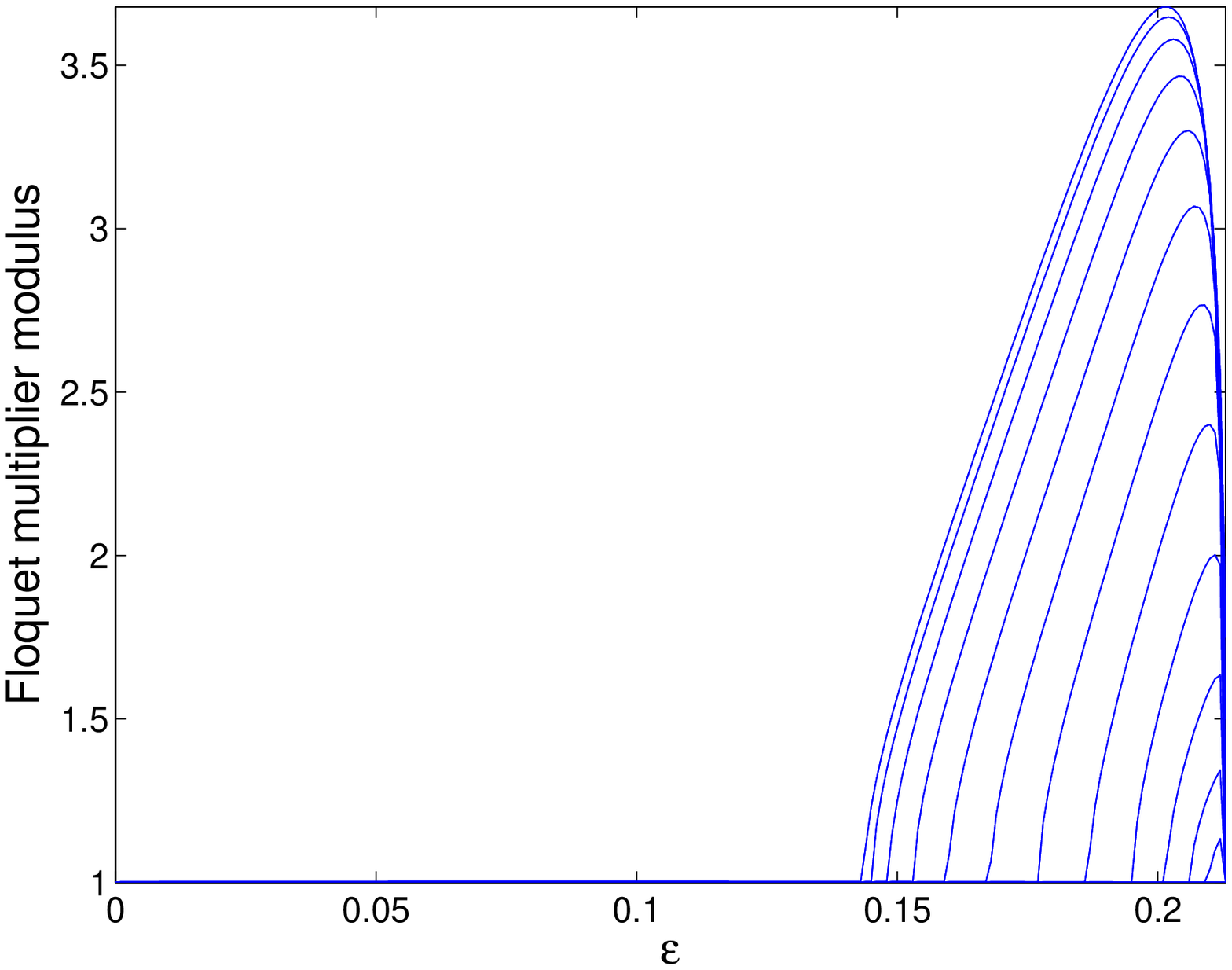} \\
\end{tabular}
\caption{Dependence of the Floquet multipliers with respect to $\ee$ for phonobreathers in a Morse potential. Parameters are: $\wb=0.68$ and $\a=4\pi/9$ (top panels), $\wb=0.8$ and $\a=4\pi/9$ (middle panels) and $\wb=0.8$ and $\a=2\pi/3$ (bottom panels).} \label{fig:stabMorse}
\end{center}
\end{figure}

Fig. \ref{fig:planes} depicts the existence and stability ranges
for phonobreathers with the above mentioned parameters whereas
Fig. \ref{fig:stabMorse} shows the dependence with respect  to
$\ee$ of the arguments and moduli of Floquet multipliers for
several examples. From these figures, it can be deduced that
instabilities and stabilities arise by means of cascades of
Neimark--Sacker bifurcations (also known as Krein crunches). An
analysis of the dynamics shows that those instabilities lead to
the disappearance of the phase torsion and the time and spatial
periodicities of the phonobreathers.

\begin{figure}
\begin{center}
\begin{tabular}{cc}
    \includegraphics[width=\middlefig]{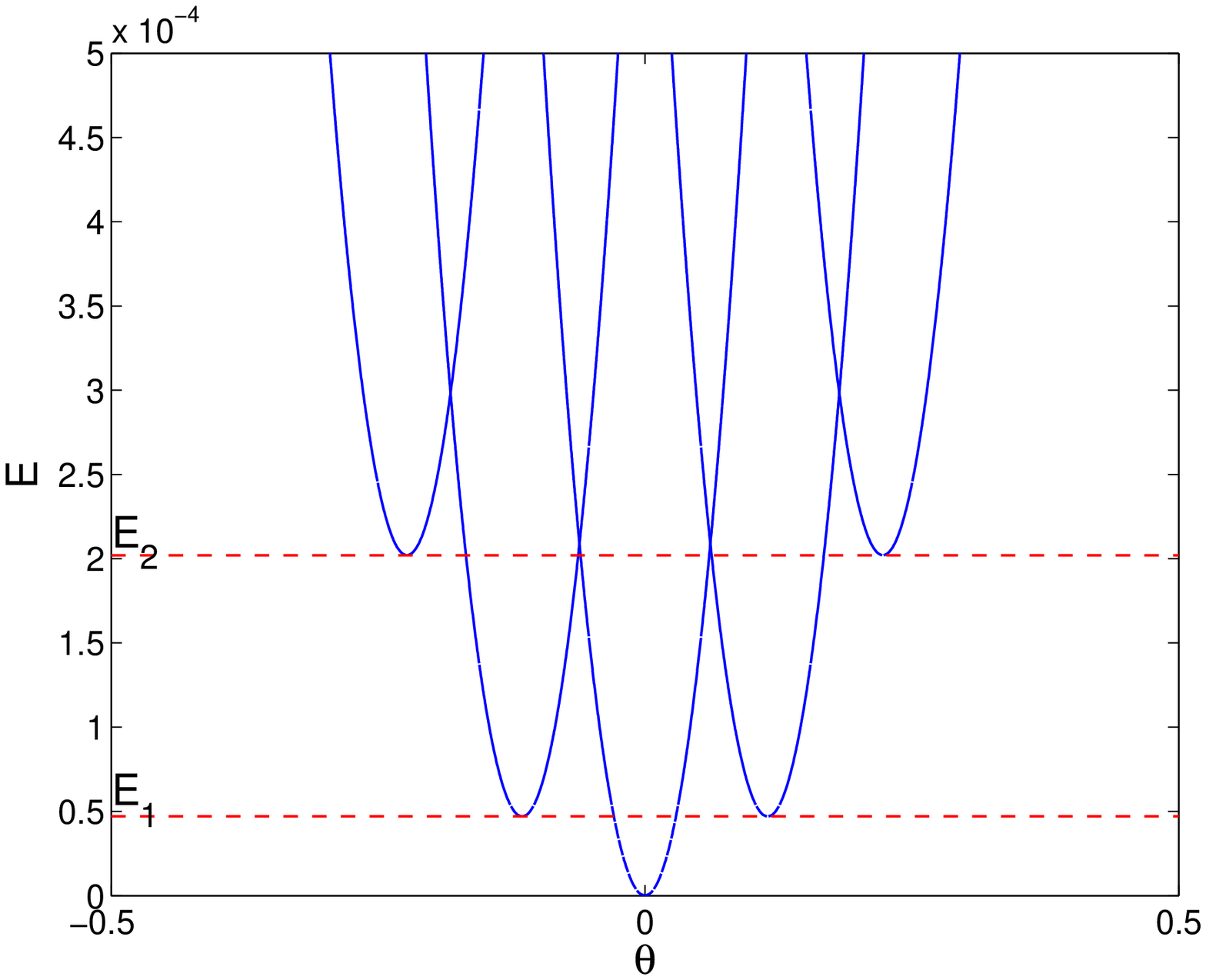} &
    \includegraphics[width=\middlefig]{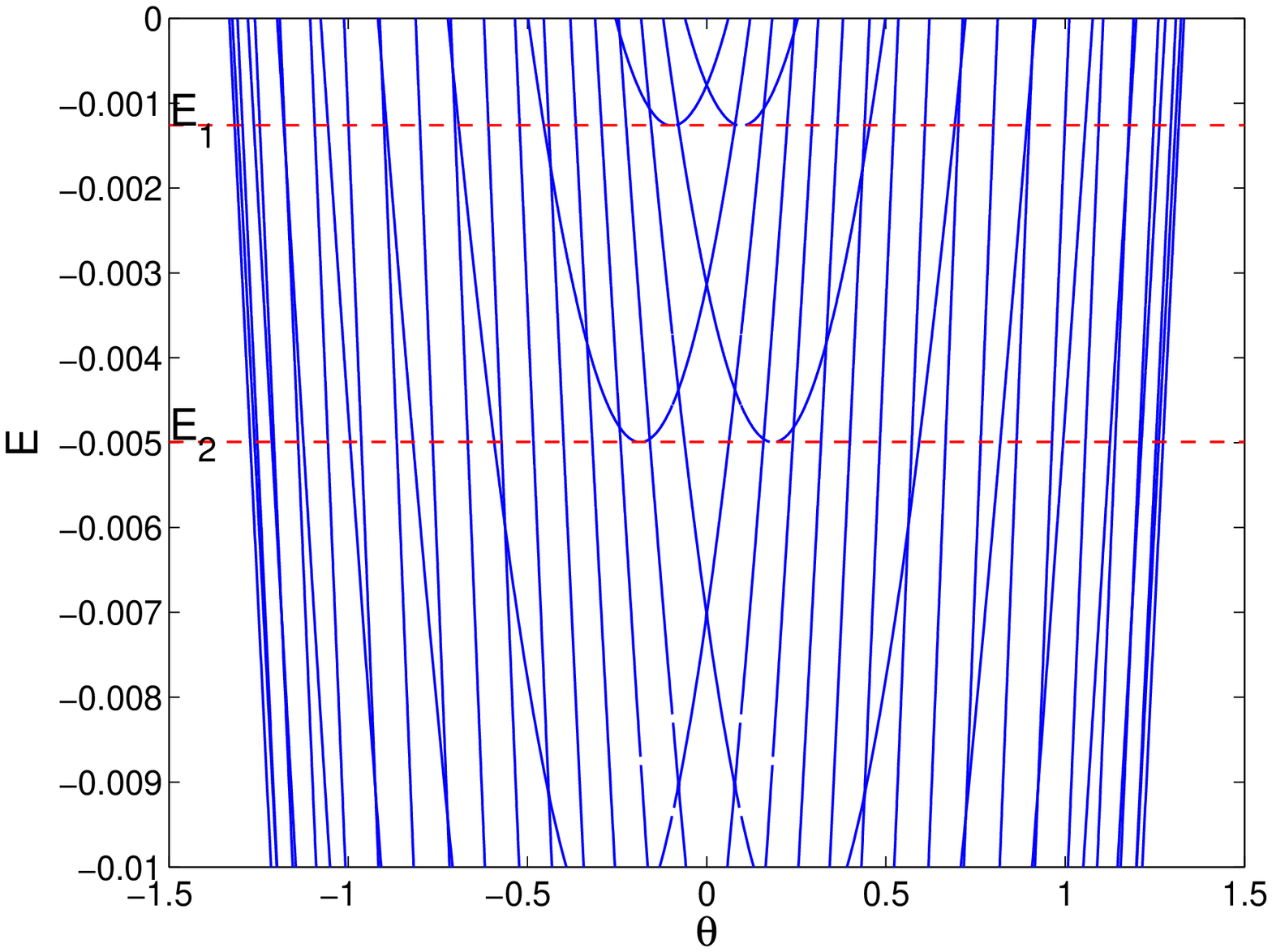} \\
\end{tabular}
\caption{Bands for a phonobreather with $\ee=0.05$, $\wb=0.8$ and
$\alpha=4\pi/9$ (left) and $\alpha=2\pi/3$ (right) and a Morse
potential. The number of particles is $N=27$. }
\label{fig:bandsMorse}
\end{center}
\end{figure}

\begin{figure}
\begin{center}
\begin{tabular}{cc}
    \includegraphics[width=\middlefig]{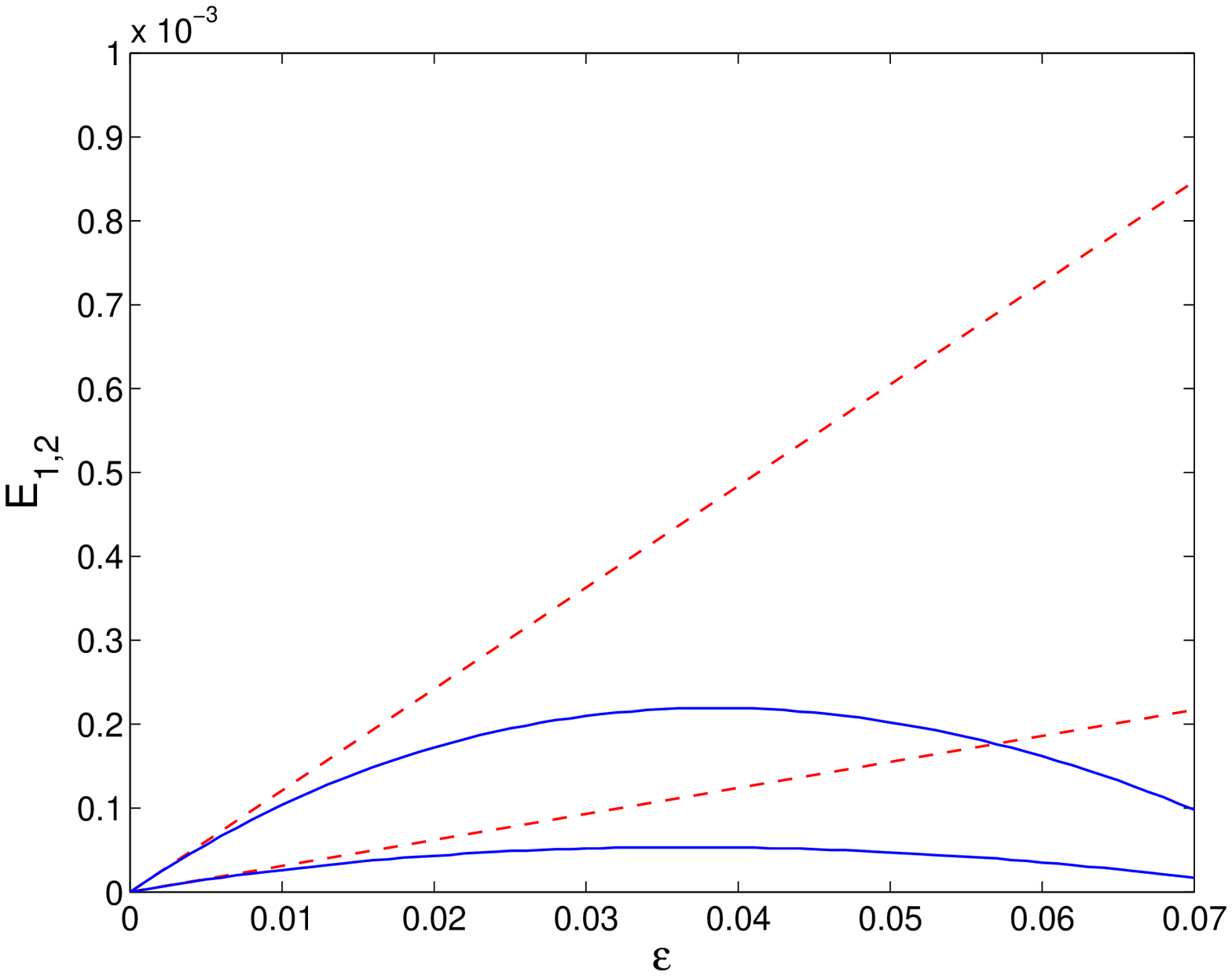} &
    \includegraphics[width=\middlefig]{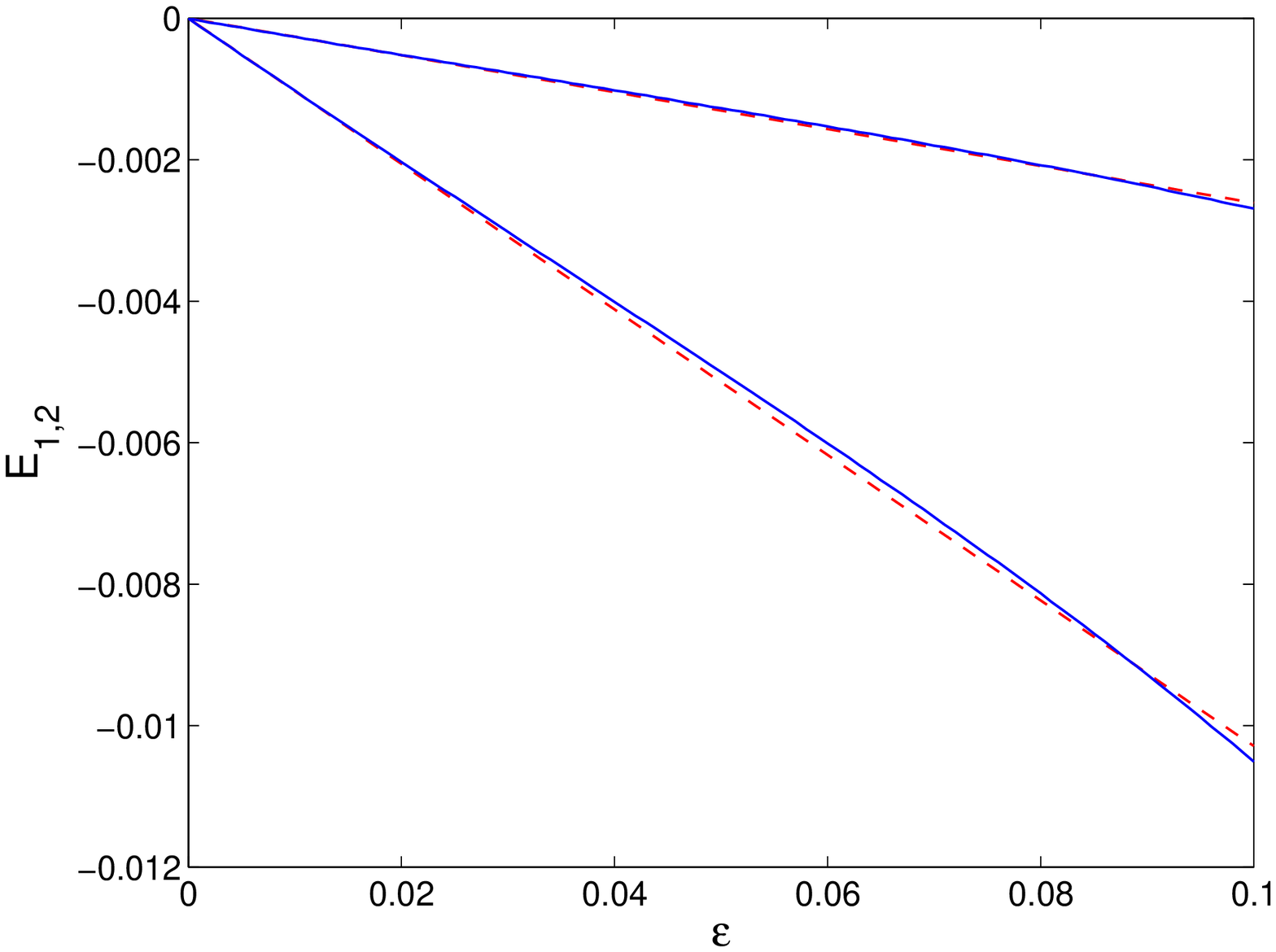} \\
\end{tabular}
\caption{Dependence of the location of the minima of the two bands
closest to $E=0$ (represented by $E_1$ and $E_2$) with respect to
$\ee$ for phonobreathers in a Morse potential with
$\alpha=4\pi/9$ (left) and $\alpha=2\pi/3$ (right). In both cases,
$\wb=0.8$. Dashed lines correspond to the MST predictions and full
lines to the numerical values. Solutions at left, where the
discrepancy is high, are unstable for small but finite coupling.}
\label{fig:bandspredMorse}
\end{center}
\end{figure}

We have checked the analytical predictions of MST regarding to
Aubry's bands displacement. To this end, we show in Fig.
\ref{fig:bandsMorse} the bands for a phonobreather with small
$\ee$ whereas Fig. \ref{fig:bandspredMorse} shows the comparison
of the predicted vertical displacement given by the MST with the
numerical value. We can observe that, contrary to time-reversible
multibreathers (where the displacement is purely vertical), there
is a diagonal movement of the bands. In addition, there is a
horizontal splitting of the bands that preserves the mirror
symmetry of the bands with respect to $E$-axis. It is also found
an excellent agreement with the analytical predictions even for
high coupling when $\alpha=2\pi/3$ (the displacement is almost
lineal), whereas for $\alpha=4\pi/9$ there is a serious
discrepancy as the bands displacement is non-monotonic and
non-linear.

\subsection{Hard $\phi^4$ potential}

\begin{figure}
\begin{center}
\begin{tabular}{cc}
    \includegraphics[width=\middlefig]{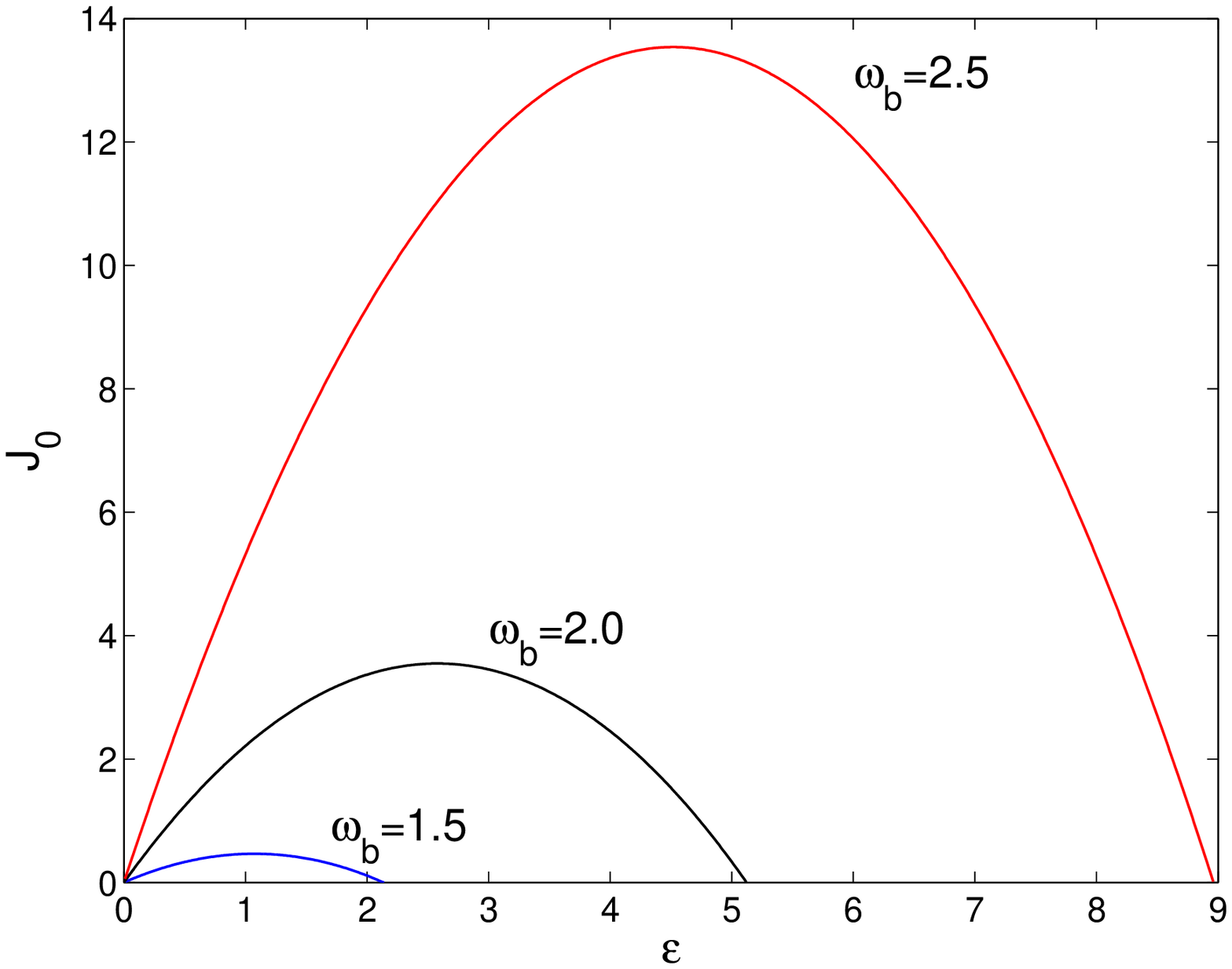} &
    \includegraphics[width=\middlefig]{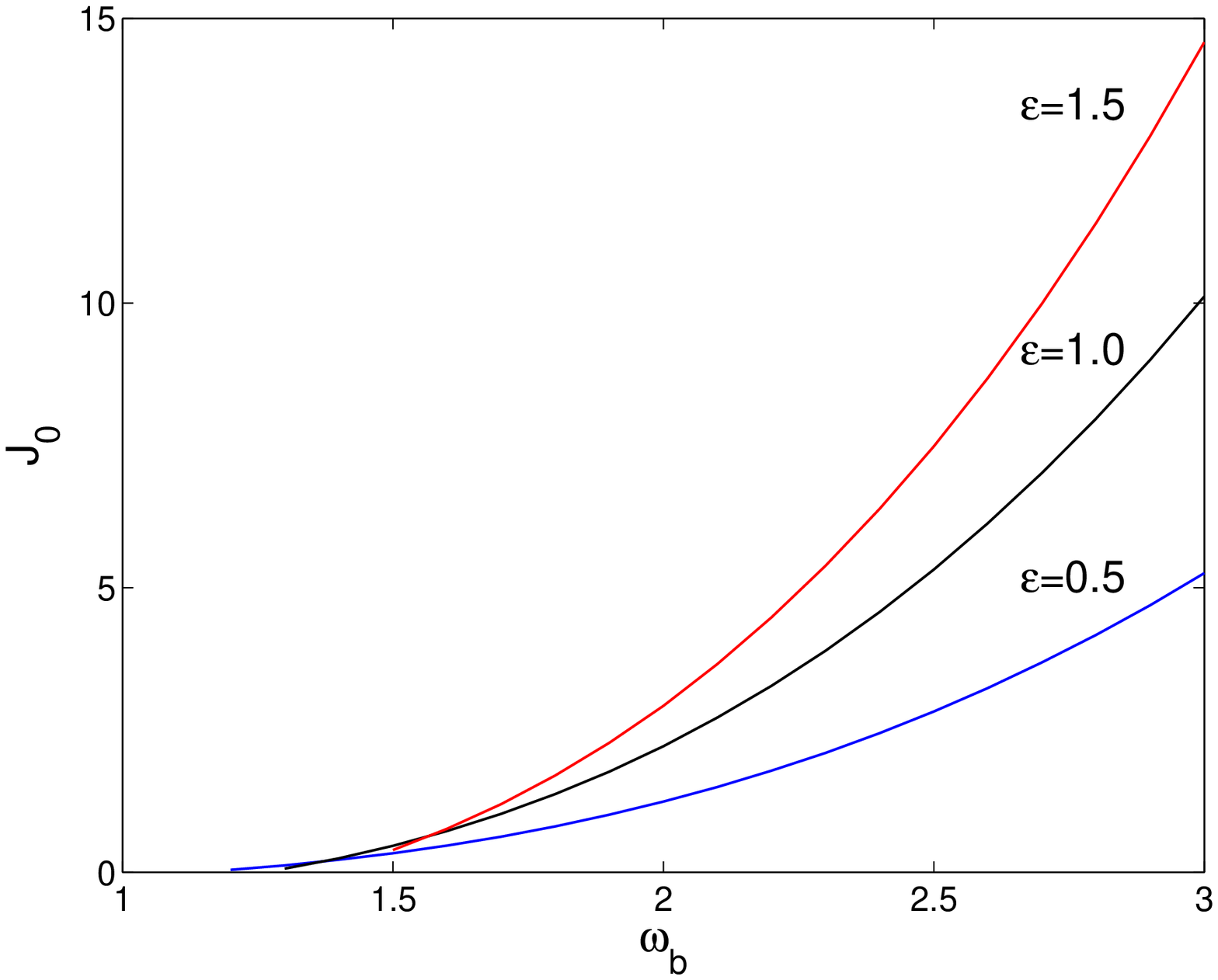} \\
    \includegraphics[width=\middlefig]{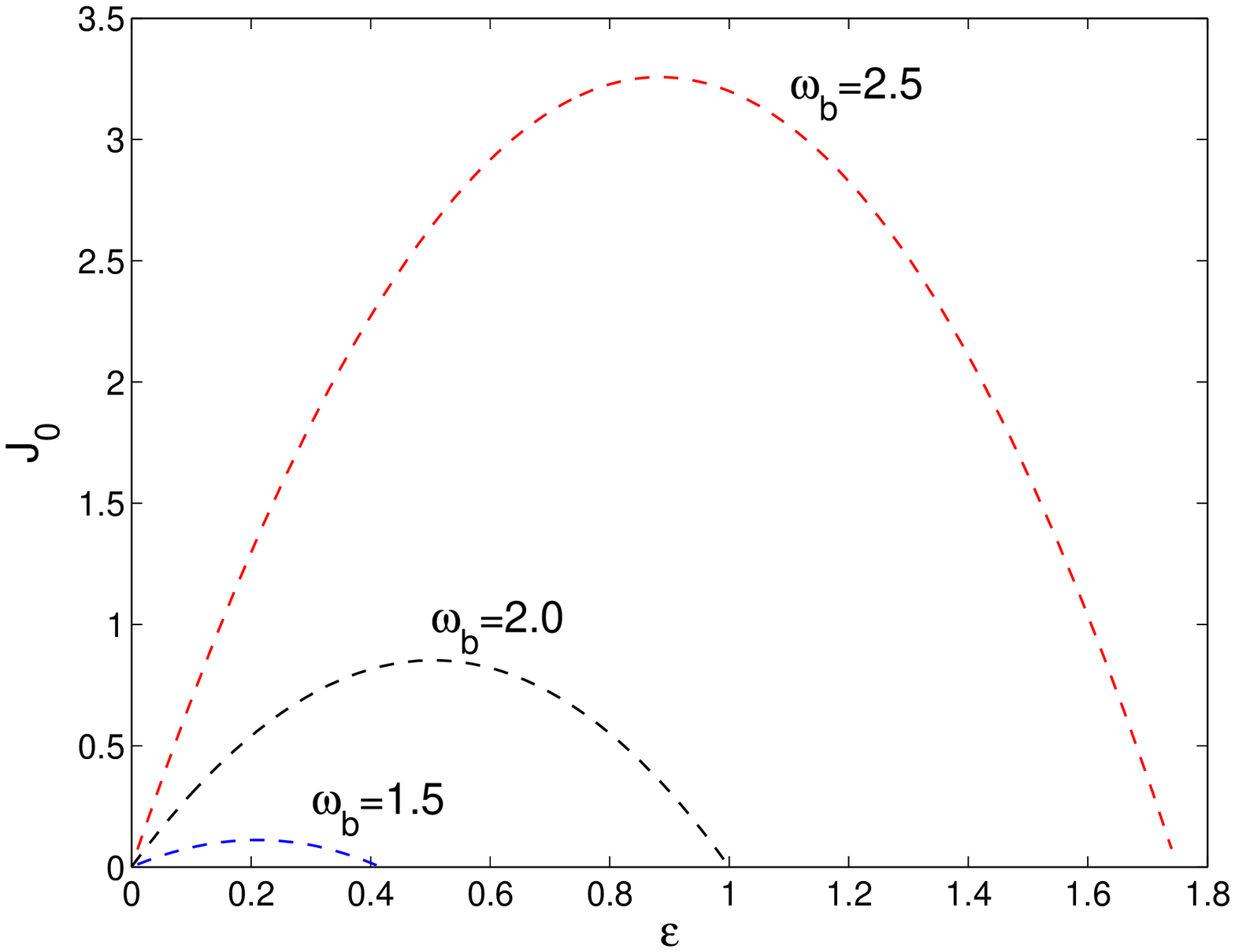} &
    \includegraphics[width=\middlefig]{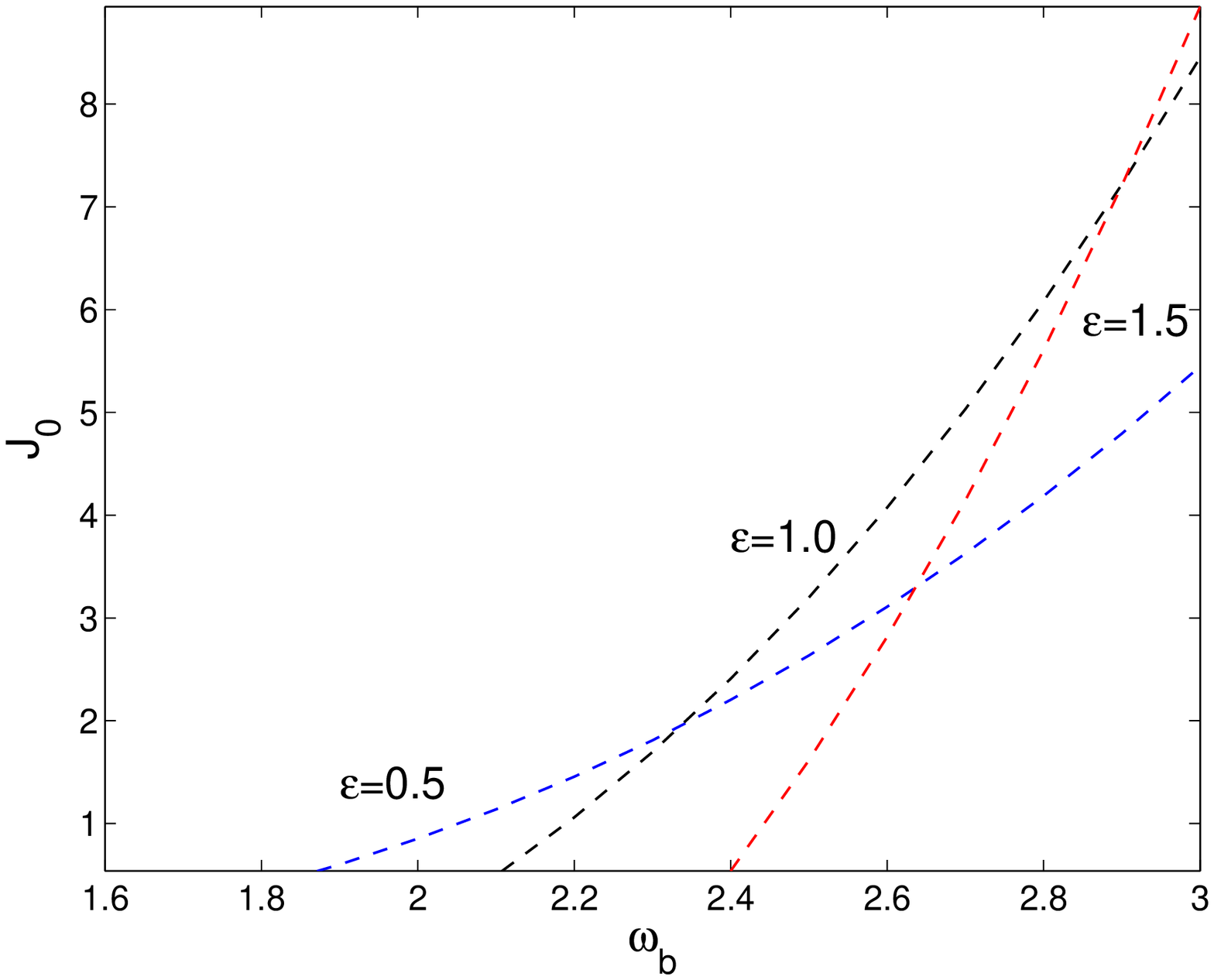} \\
\end{tabular}
\caption{Flux between nearest-neighbours for phonobreathers in a $\phi^4$ potential with $\a=\pi/4$ (top panels) and $\a=3\pi/2$ (bottom panels). Dashed lines represent to unstable solutions. Note that the flux tends to zero when the continuation in $\ee$ fails.} \label{fig:flux2}
\end{center}
\end{figure}

The analysis of the previous case can also be done for the hard
$\phi^4$ potential. We start by showing in Fig.~\ref{fig:flux2}
the properties of the flux for $\a=\pi/4$ and $\a=3\pi/2$; the
qualitative behaviour displayed in the figure is generic for any
value of $\a$. Contrary to the Morse case, the flux has a
non-monotonical behavior, increasing with $\ee$ for small coupling
and decreasing for large coupling; additionally, the flux is zero
for zero coupling and for a critical value of $\ee$. However,
the flux increases with the frequency. As in the Morse case, this
behaviour allowed us to calculate the critical coupling as a
function of the frequency, as shown in \ref{app:crit}.

\begin{figure}
\begin{center}
\begin{tabular}{cc}
    \includegraphics[width=\middlefig]{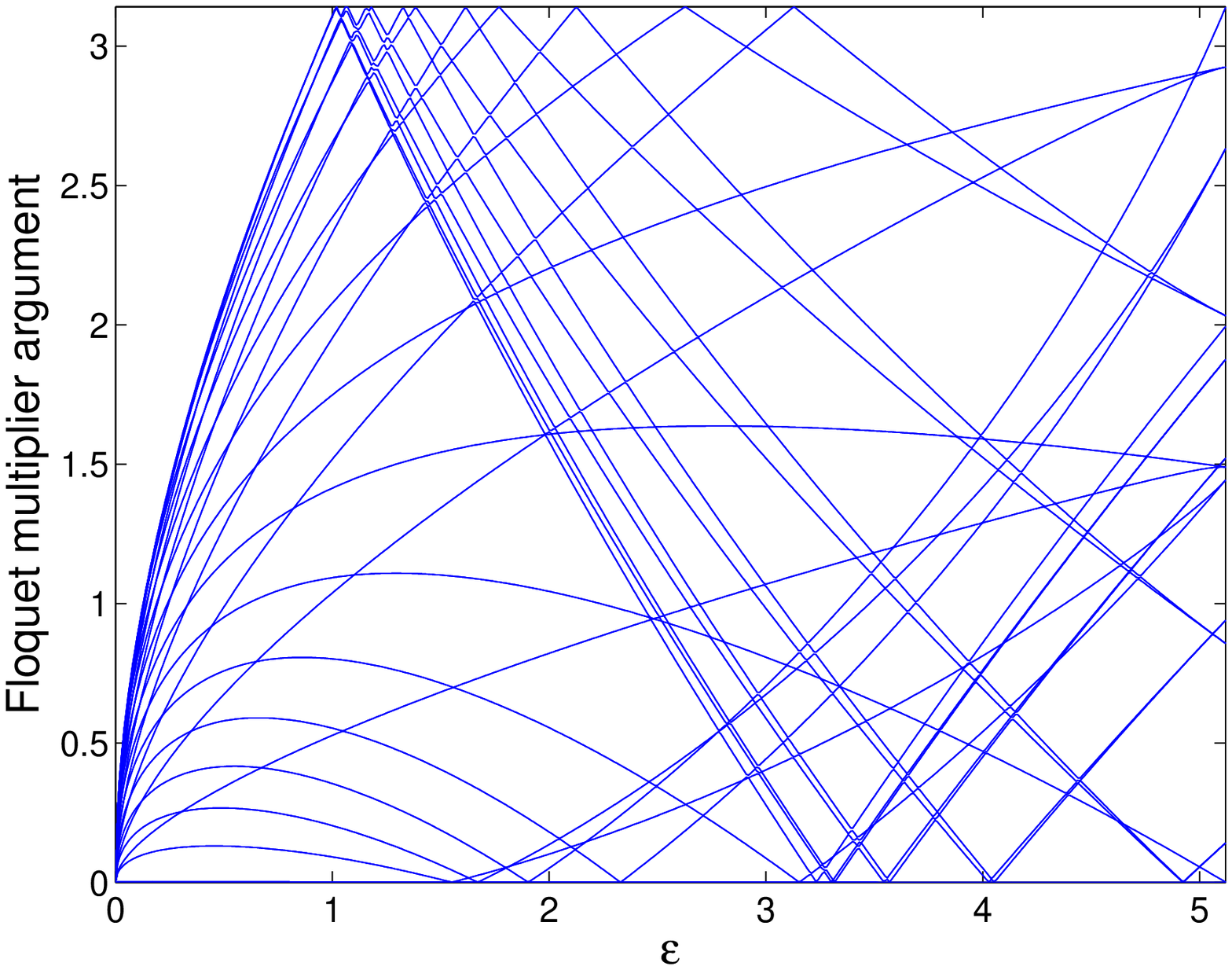} &
    \includegraphics[width=\middlefig]{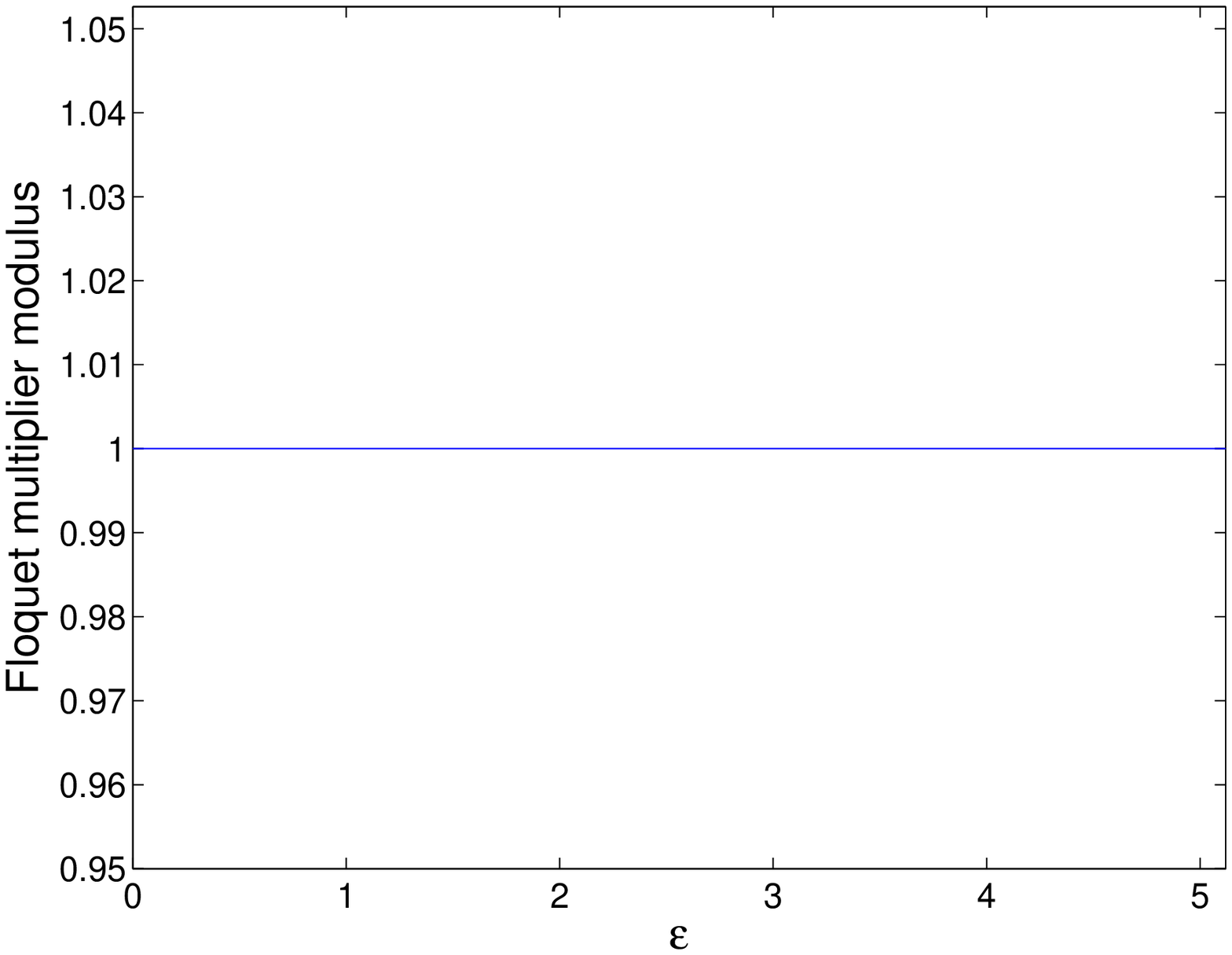} \\
    \includegraphics[width=\middlefig]{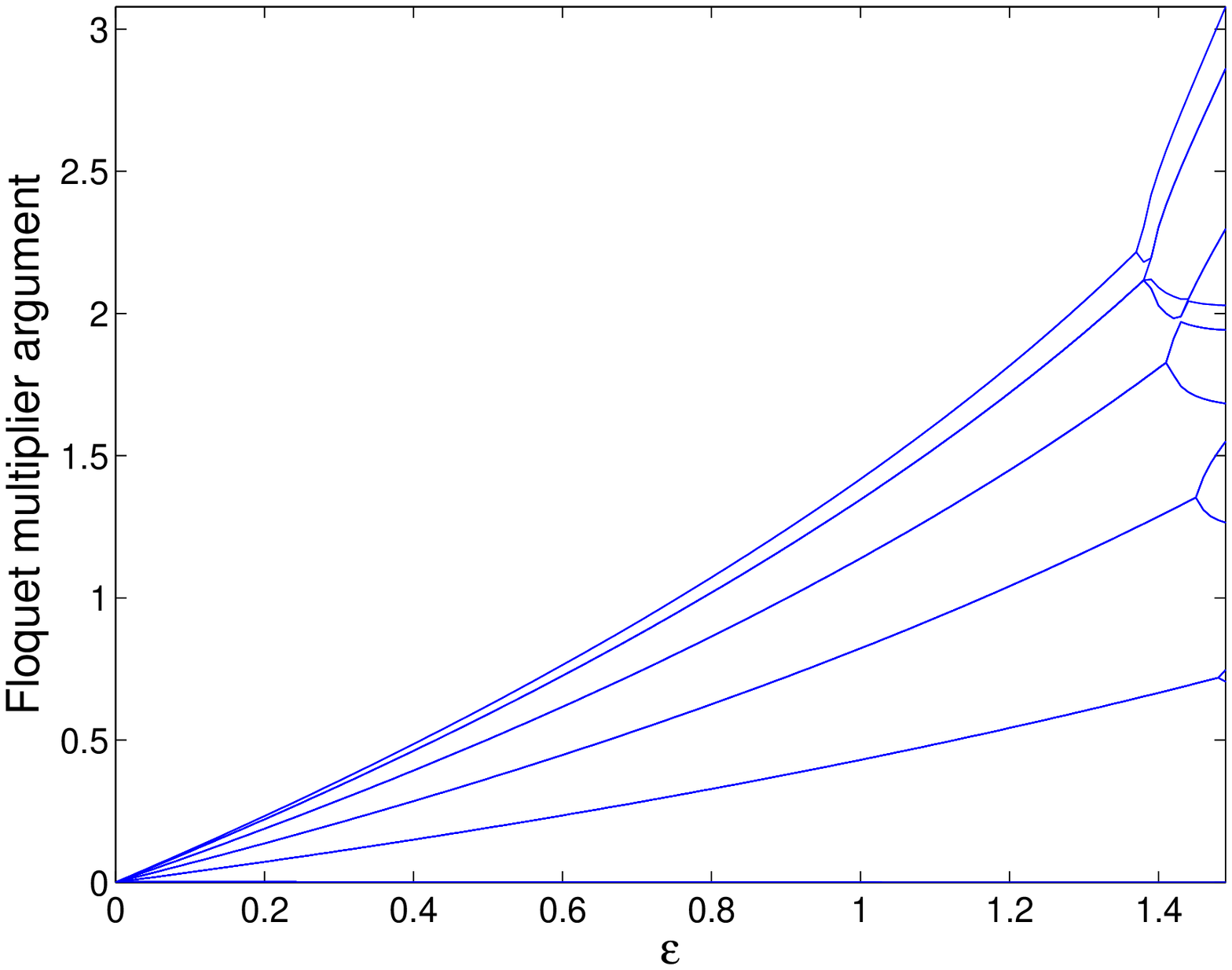} &
    \includegraphics[width=\middlefig]{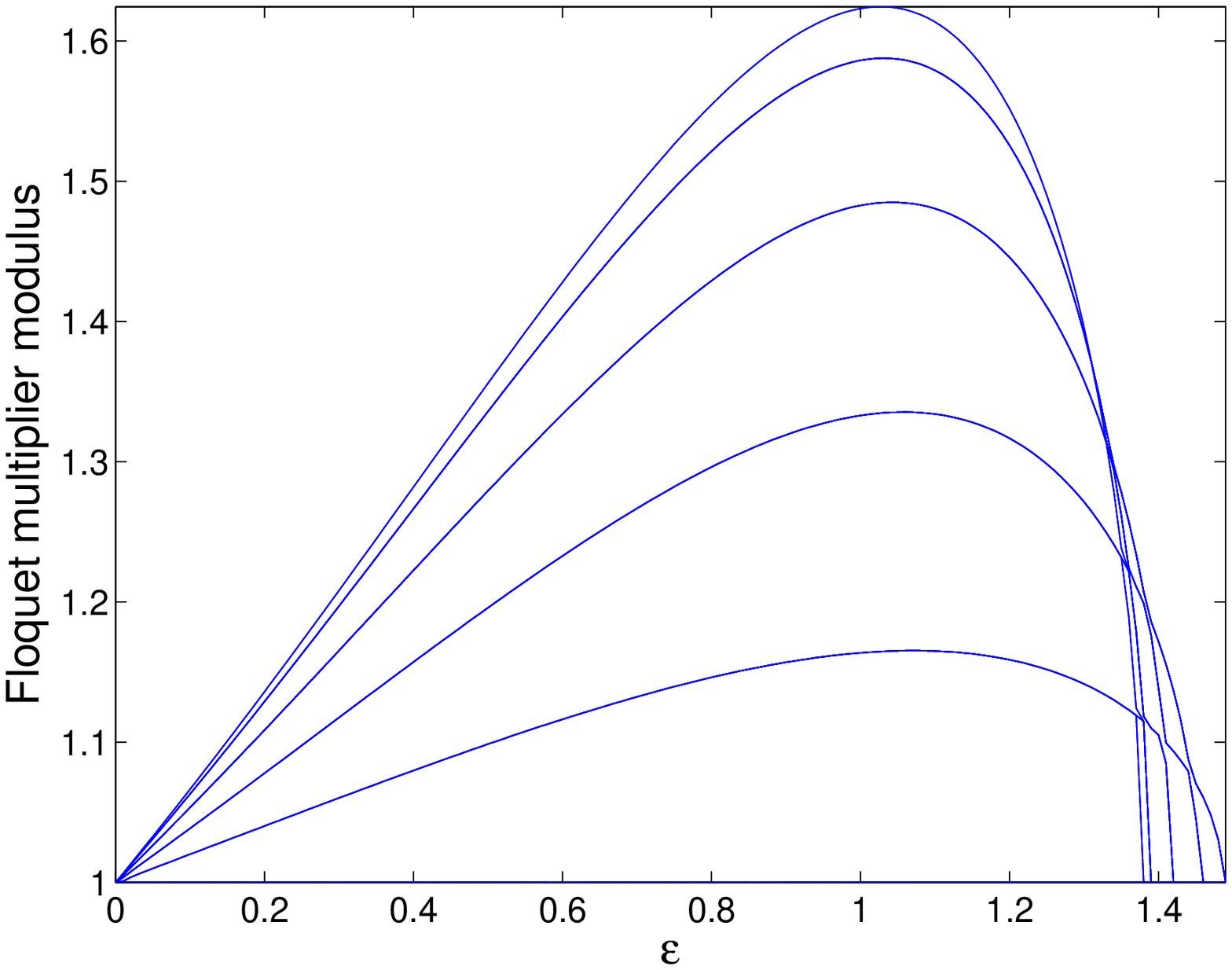} \\
    \includegraphics[width=\middlefig]{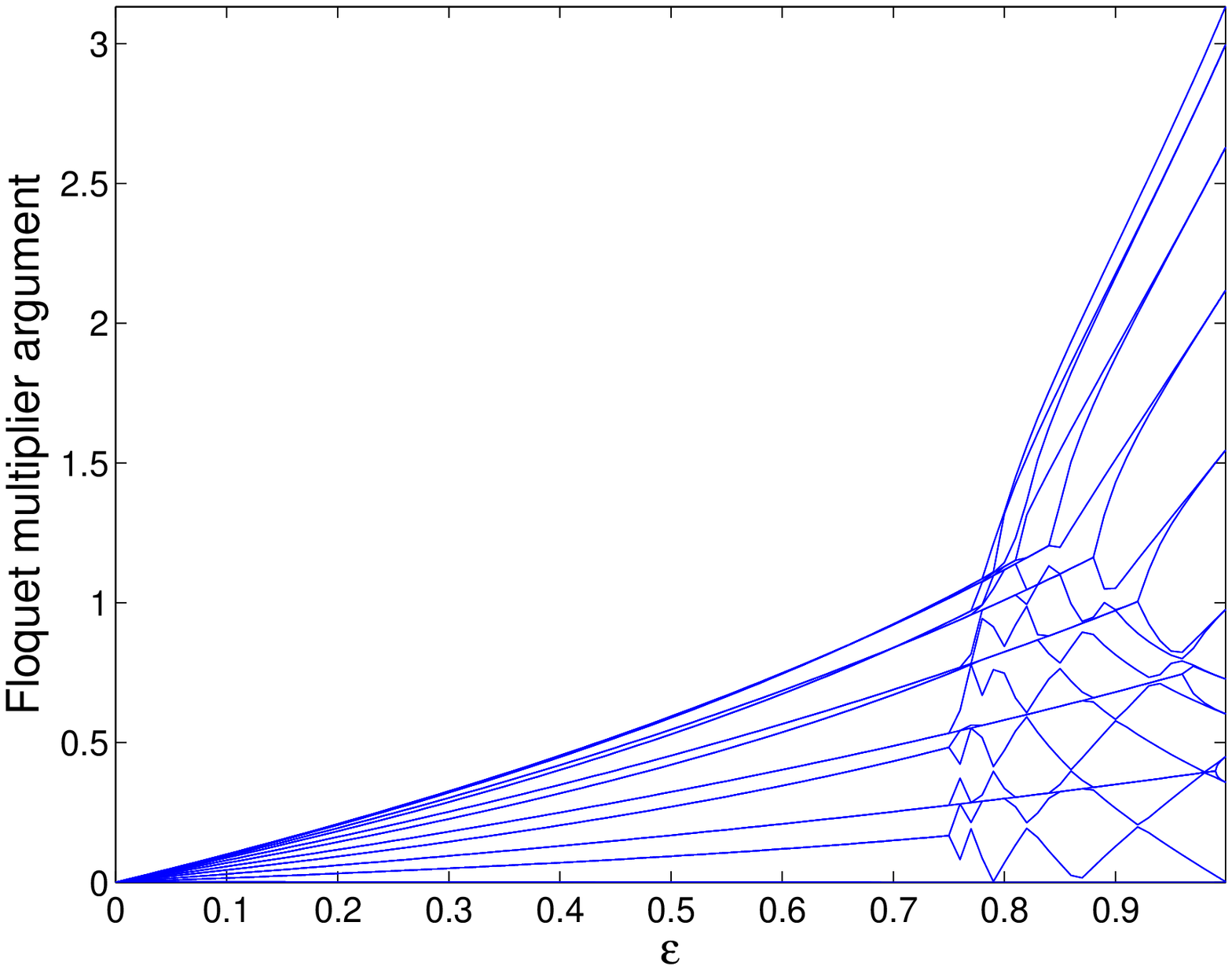} &
    \includegraphics[width=\middlefig]{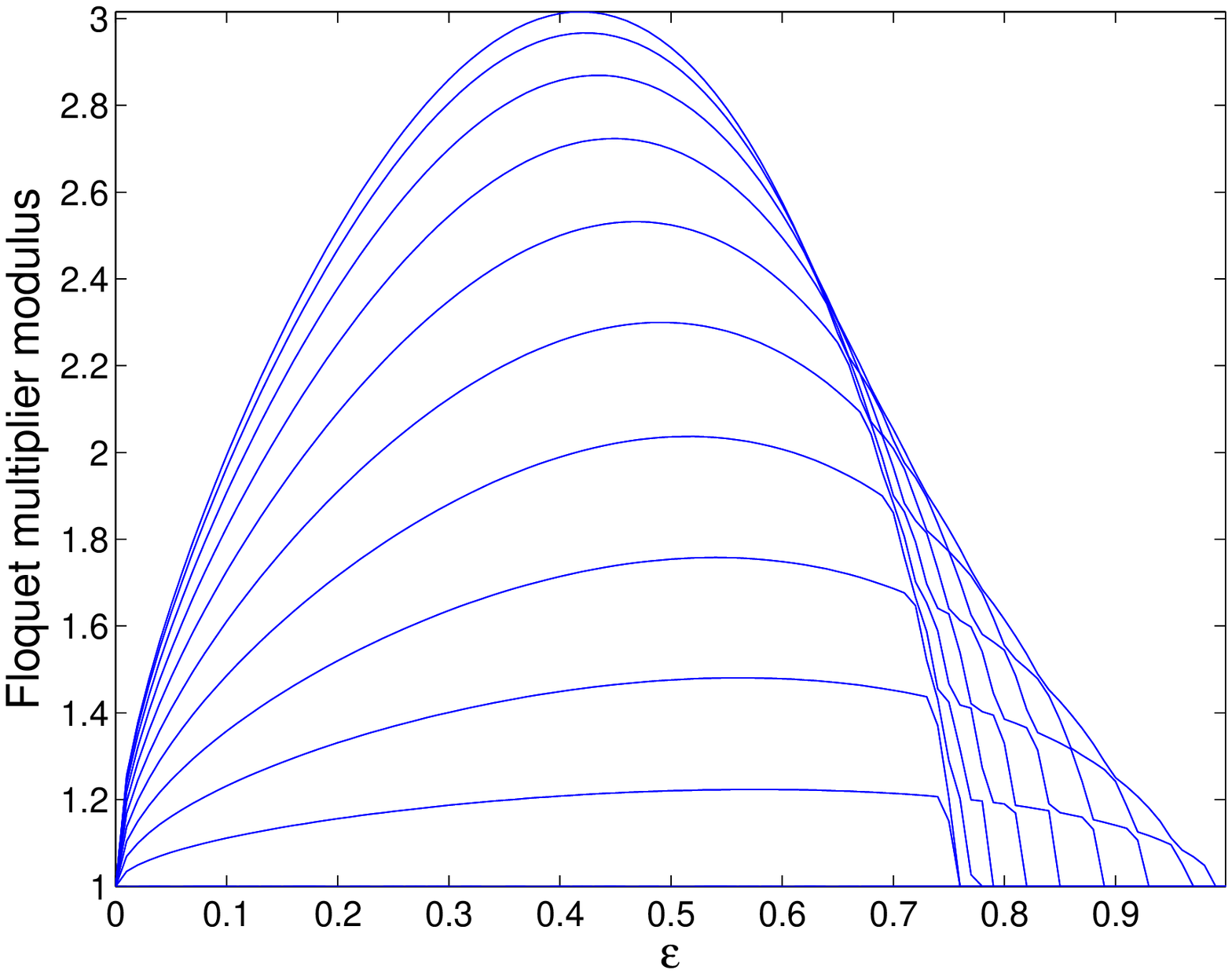} \\
\end{tabular}
\caption{Dependence of the Floquet multipliers with respect to $\ee$ for phonobreathers in a $\phi^4$ potential. Parameters are: $\wb=2$ and $\a=2\pi/3$ (top panels), $\wb=2$ and $\a=\pi/2$ (middle panels) and $\wb=2$ and $\a=\pi/4$ (bottom panels).} \label{fig:stabphi4}
\end{center}
\end{figure}

Let us show the results regarding to the stability. Analogously to
the time-reversible breathers, there exist no stability exchange
bifurcations and the stability for any arbitrary $\ee$ is the same
as for small coupling except for $\a=\pi/2$. Thus, for $\a>\pi/2$ ($\a<\pi/2$),
phonobreathers stability is the same as in the $\a=\pi$ ($\a=0$)
case. The only unknown appears at $\a=\pi/2$, where the
multibreathers stability theorem fails. In that case, numerical
calculations demonstrate that the stability range depends both on $\ee$ and
$\wb$. Fig. \ref{fig:stabphi4} displays the dependence of the
Floquet multipliers with respect to $\ee$ for $\alpha=2\pi/3$,
$\alpha=\pi/2$ and $\alpha=\pi/4$. In all cases, for a given
$\wb$, there is a critical value of $\ee$ where the continuation
fails. At this value the flux vanishes, contrary to the Morse
potential, for which the flux diverges. This critical value of
$\ee$ has been calculated in Eq. \ref{eq:critphi4}:

\begin{equation}
    \ee_c=\frac{\wb^2-1}{4\sin^2(\alpha/2)}
\end{equation}

\begin{figure}
\begin{center}
\begin{tabular}{cc}
    \includegraphics[width=\middlefig]{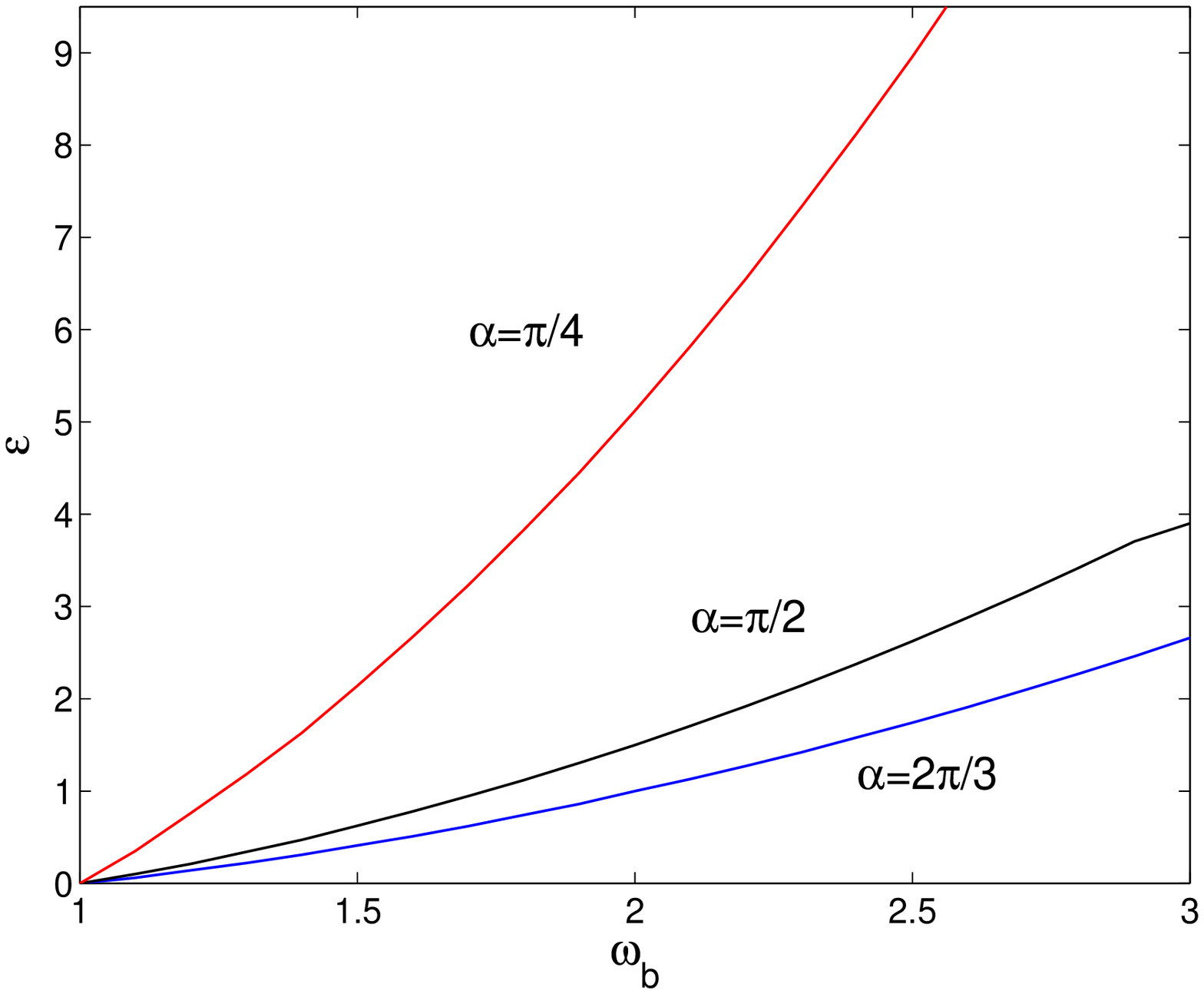}&
    \includegraphics[width=\middlefig]{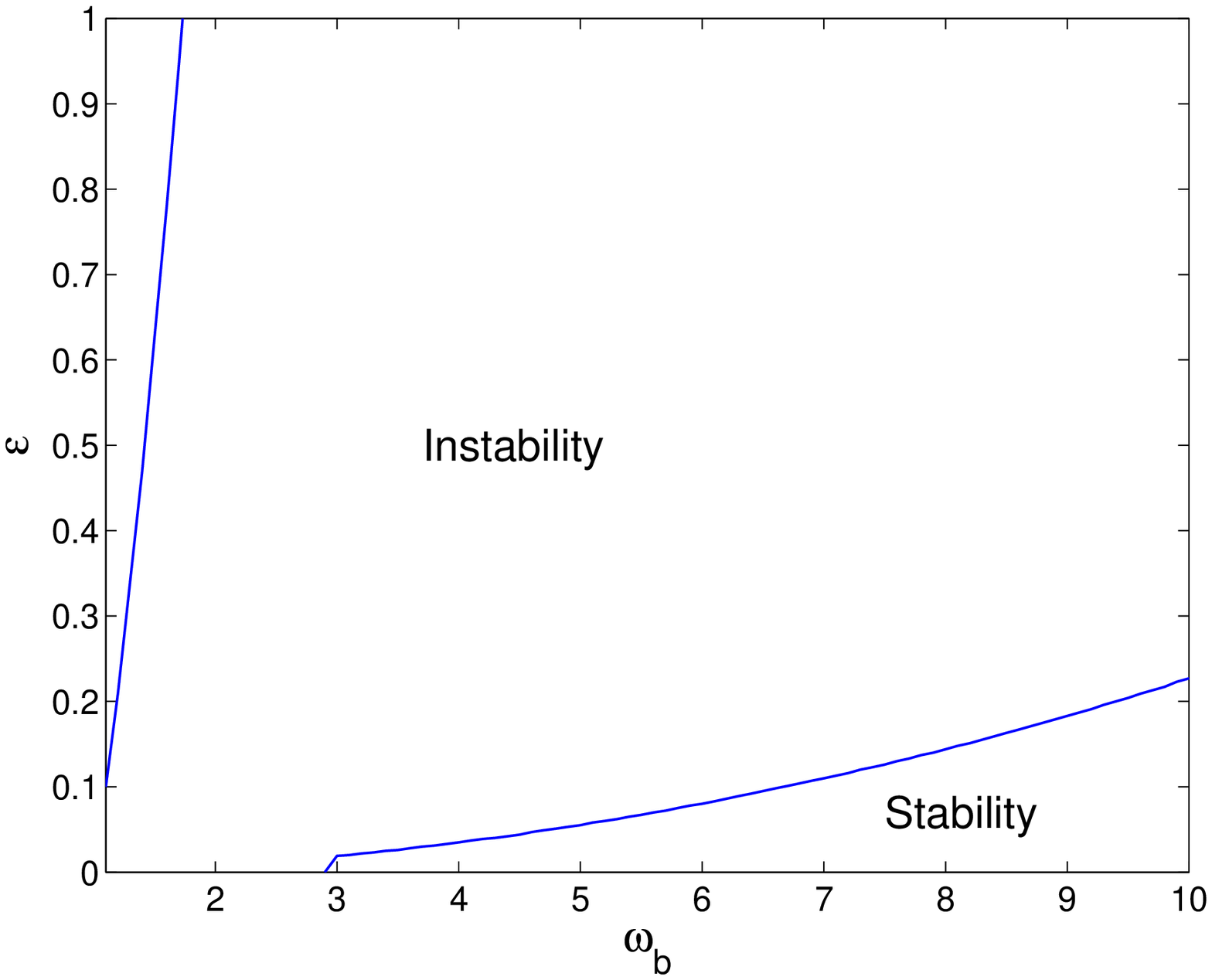}\\
\end{tabular}
\caption{(Left) $\ee$-$\wb$ plane showing the existence of phonobreathers $\phi^4$ potential. (Right) shows the existence and stability range for phonobreathers with $\alpha=\pi/2$.} \label{fig:planes2}
\end{center}
\end{figure}

Fig. \ref{fig:planes2} shows the existence range for those values of $\alpha$. It can be deduced that the existence range is enlarged when $\alpha$ decreases. Besides, for $\alpha=\pi/2$ it is observed that the stability properties depends on the frequency and coupling and that phonobreathers are always unstable for $\wb\leq3$.

\begin{figure}
\begin{center}
\begin{tabular}{cc}
    \includegraphics[width=\middlefig]{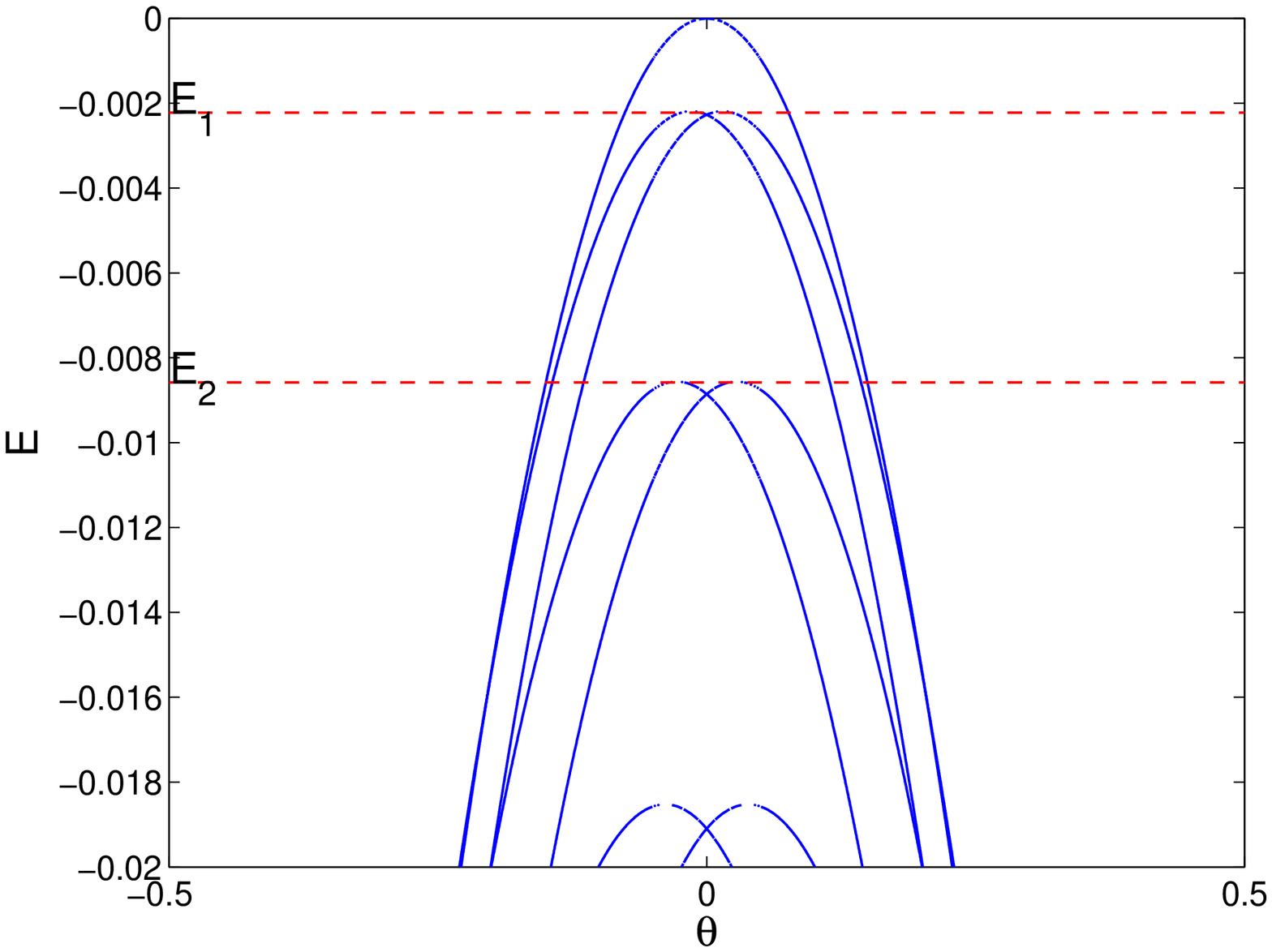} &
    \includegraphics[width=\middlefig]{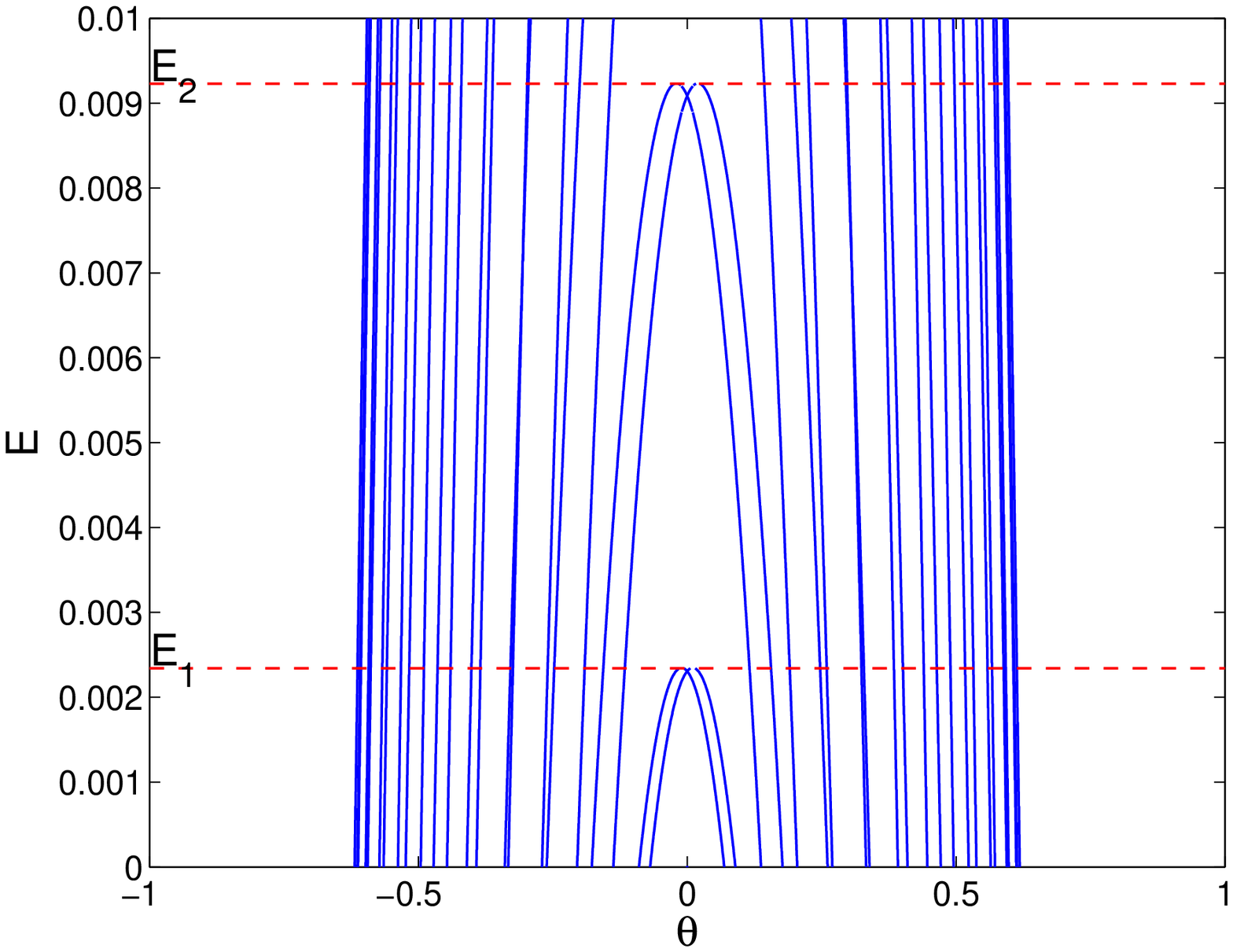} \\
\end{tabular}
\caption{Bands for a phonobreather with $\ee=0.05$, $\wb=2$ and $\alpha=2\pi/3$ (left) and $\alpha=\pi/4$ (right) and a $\phi^4$ potential. The number of particles is $N=21$ (left) and $N=24$ (right).} \label{fig:bandsphi4}
\end{center}
\end{figure}

\begin{figure}
\begin{center}
\begin{tabular}{cc}
    \includegraphics[width=\middlefig]{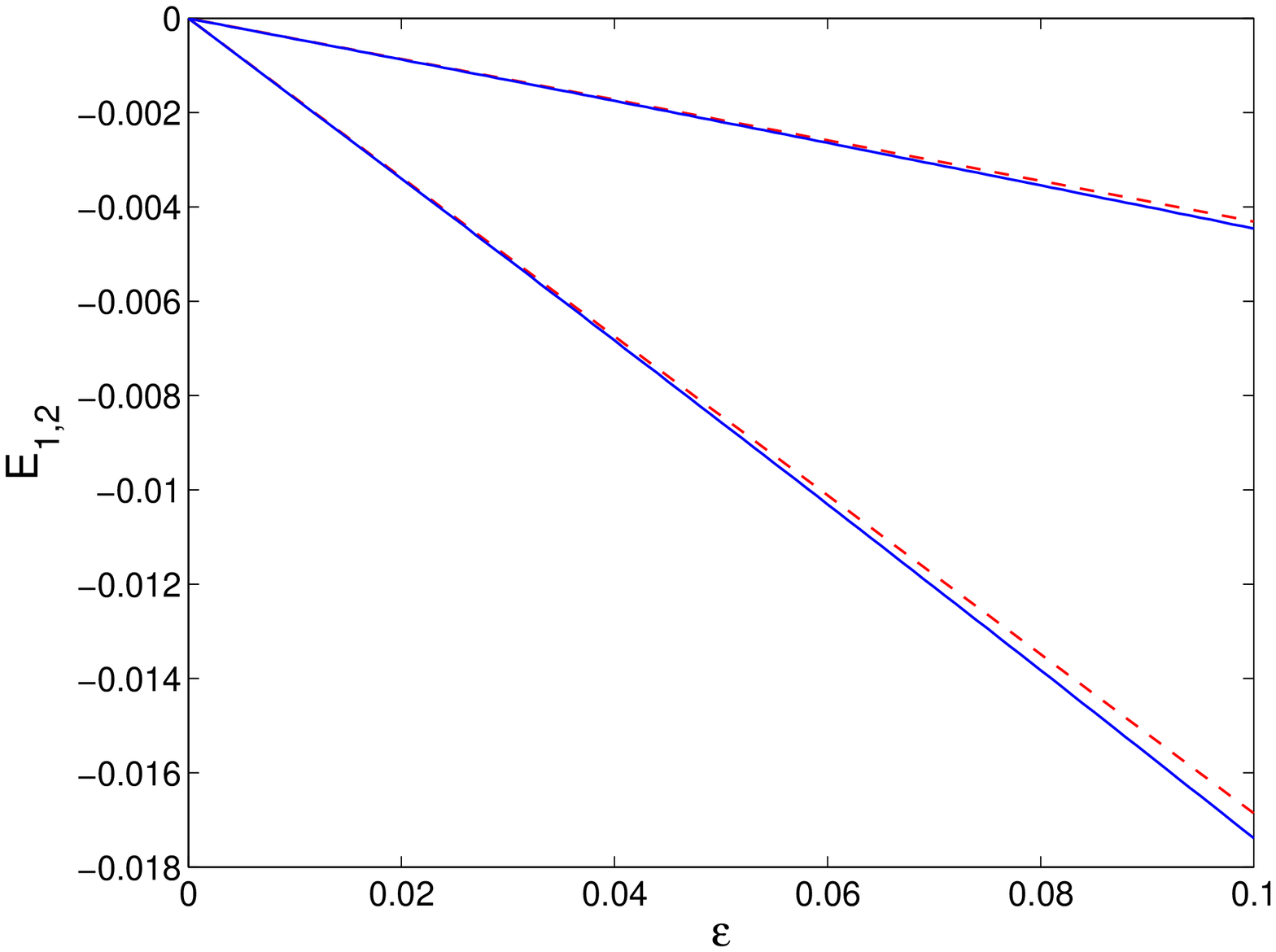} &
    \includegraphics[width=\middlefig]{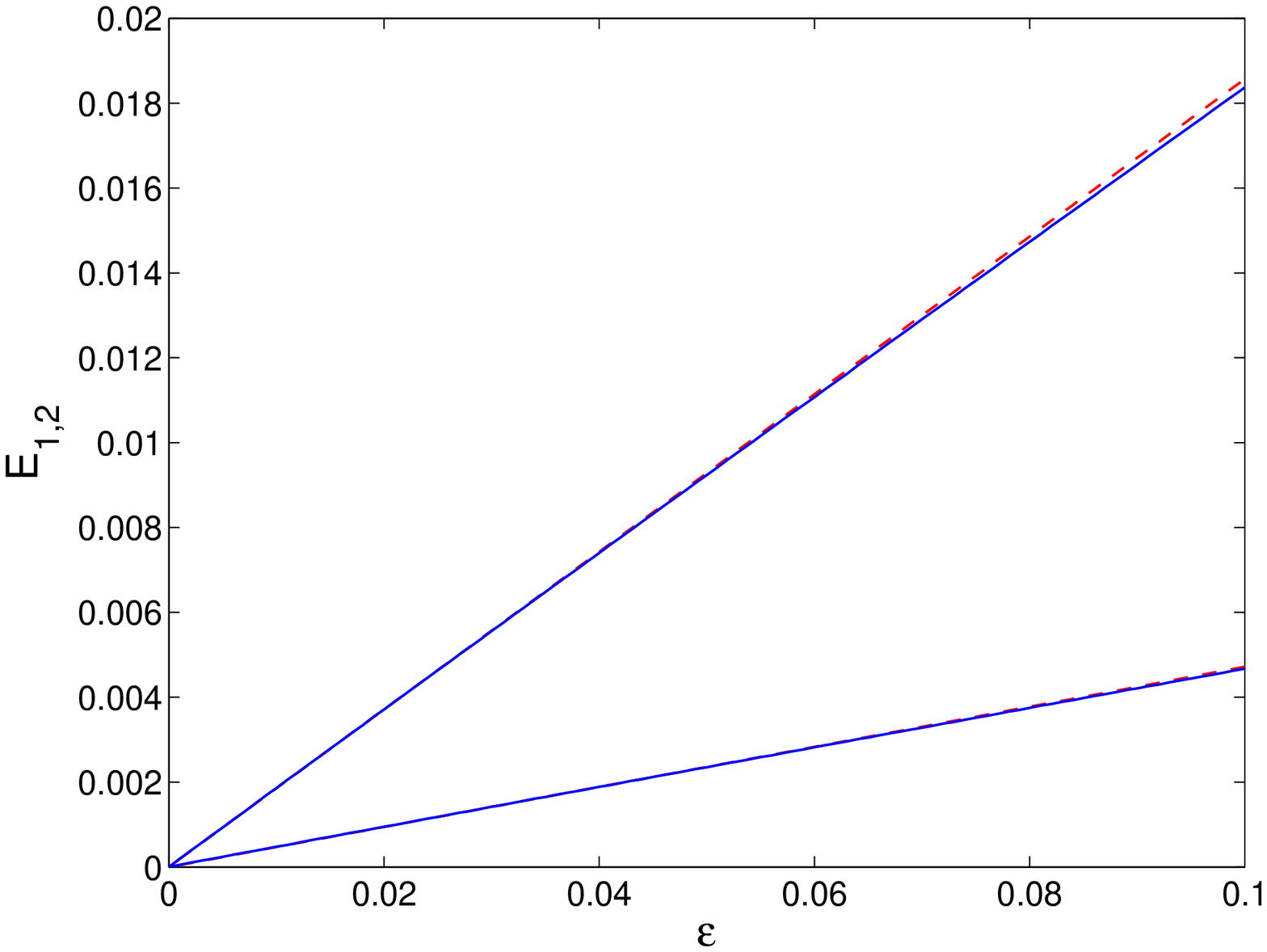} \\
\end{tabular}
\caption{Dependence of the location of the maxima of the two bands closest to $E=0$ (represented by $E_1$ and $E_2$) with respect to $\ee$ for phonobreathers in a $\phi^4$  with $\alpha=2\pi/3$, $N=21$ (left) and $\alpha=\pi/4$, $N=24$ (right). In both cases, $\wb=0.8$. Dashed lines correspond to the MST predictions and full lines to the numerical values. Dashed lines correspond to the MST predictions and full lines to the numerical values.} \label{fig:bandspredphi4}
\end{center}
\end{figure}

\begin{figure}
\begin{center}
\begin{tabular}{cc}
    \includegraphics[width=\middlefig]{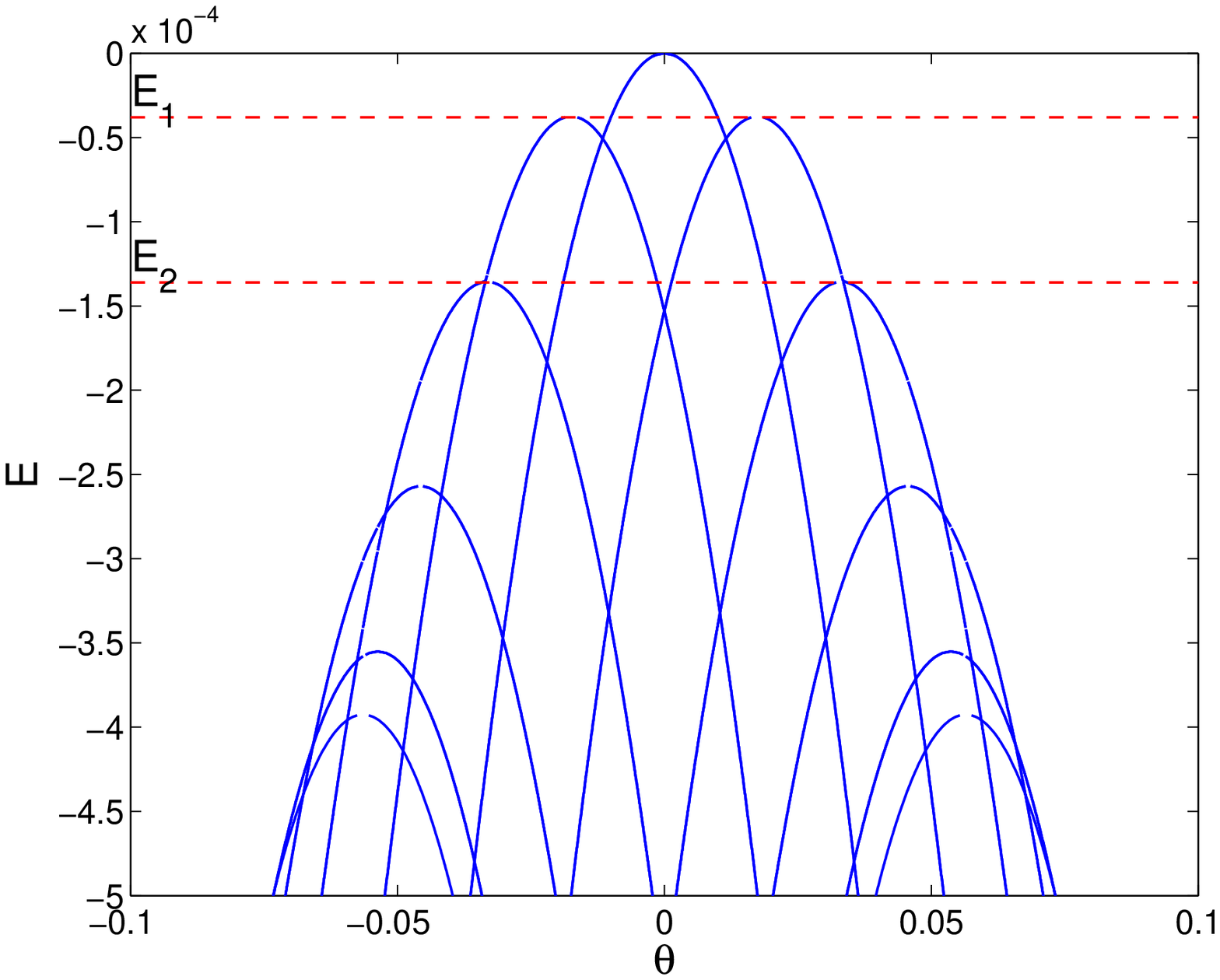} &
    \includegraphics[width=\middlefig]{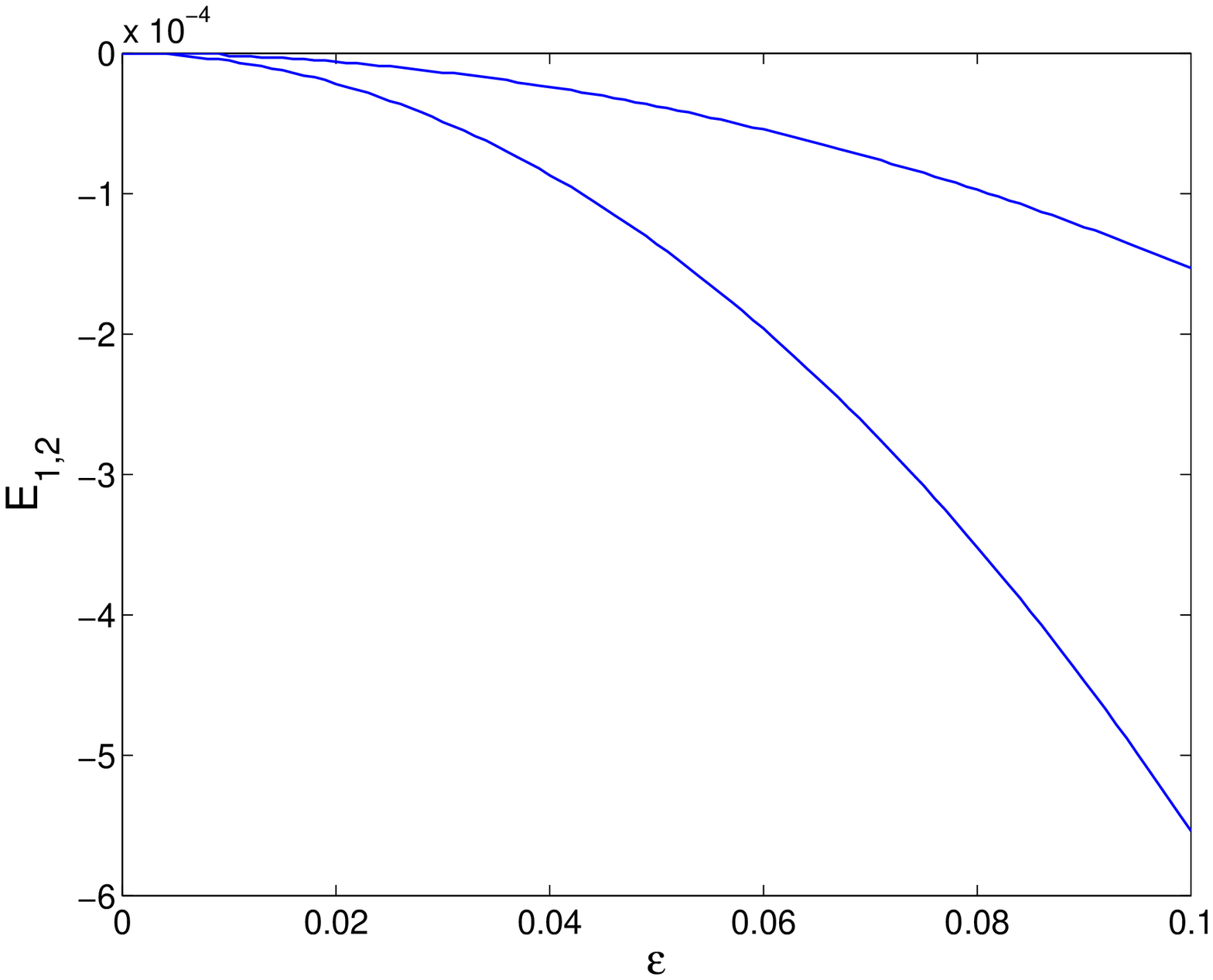} \\
\end{tabular}
\caption{Phonobreather in a $\phi^4$ potential with $\wb=2$, $\a=\pi/2$ and $N=20$. (Left) Band diagram for $\ee=0.05$. (Right) Dependence of the location of the maxima of the two bands closest to $E=0$ with respect to $\ee$.} \label{fig:bandspihalf}
\end{center}
\end{figure}

Fig. \ref{fig:bandsphi4} and Fig. \ref{fig:bandspredphi4} show, respectively, examples of bands and the MST predictions, for $\alpha=2\pi/3$ and $\alpha=\pi/4$. There is an excellent agreement of the numerical results with the analytical predictions. Additionally, Fig. \ref{fig:bandspihalf} focuses on the $\alpha=\pi/2$ case with $\wb=2$. From the last figure, we can observe that the first and second minima of the bands adjust respectively to the parabolas $-0.015509\ee^2$ and $-0.056476\ee^2$. This fact confirms that a second order perturbation theory is needed in order to predict the stability properties for $\a=\pi/2$.

\section{Conclusions}\label{sec:conc}

The study carried out in this paper has been twofold: on the one
hand, analytical predictions on the stability of non-time
reversible phonobreathers has been established by making use of
the multibreathers stability theorem; on the other hand, a
numerical study on the existence and stability properties of those
phonobreathers has been performed.

The multibreathers stability theorem predicts that there is a
critical value of the phase difference in phonobreathers that
separates stable and unstable states. This critical value depends
on the phonobreather frequency for lattices with harmonic coupling
and Morse on-site potential whereas, for the $\phi^4$ potential,
the critical value is $\pi/2$ and does not depend on the
frequency. At this particular value of the phase, the theorem
cannot make any prediction about the (in)stability.

The analytical findings have been confirmed by means of numerical
analysis, showing in addition that the stability changes above a
given value of the coupling constant. It has also been confirmed
that, in most cases, the Aubry's band displacement predicted by
the multibreather stability theorem for small coupling, is valid
even for relatively high coupling. Apart from this, it has been
observed that phonobreathers cease to exist above a critical value
of the coupling. This value coincides with a flux divergence for
the Morse potential or a flux vanishing for the hard $\phi^4$
potential. Finally, it has been found that for the $\phi^4$
potential with $\alpha=\pi/2$, the (in)stability depends on the
frequency even for infinitesimally small coupling, being
additionally confirmed that a second-order perturbation theory is
needed in order to predict the (in)stability at this particular
value of the phase.

The above mentioned results open new perspectives for  future
work. For instance, we have started to study the properties of
vortices in square lattices, which can be pictured as
multibreathers with $\alpha=\pi/2$. It can be interesting to apply
the multibreathers stability theorem to two structures considered
by Cretegny and Aubry: (i) breather ``rivers'' (percolating
cluster) firstly and (ii) phonobreather with vacancy sites.
Finally, we have also started to study the effect  of long-range
interactions in the stability of non-time reversible structures.

\section*{Acknowledgments}

We acknowledge financial support from the MICINN project FIS2008-04848.

\clearpage

\appendix

\section{Calculation of the critical coupling}
\label{app:crit}

In this appendix we calculate the critical value of the coupling
parameter $\ee$ for the on-site potentials considered in the
paper, i.e. the Morse and the $\phi^4$ potentials, by making use
of the rotating wave approximation (RWA) (see e.g. \cite{Page}).
This method consists in considering all the Fourier coefficients
zero but the first one, as this is usually considerably larger
that the other coefficients. The RWA gives often surprisingly good
results, taken in account its simplicity.

Let us consider equation Eq.~(\ref{eq:dyn}):

\begin{equation}\label{eq:app:dyn}
    \ddot{u}_n \,+\,V'(u_n)\,+\,\ee \,(2 u_n-u_{n+1}-u_{n-1})\,=\,0,\,\quad
    n=1,\dots,N.
\end{equation}

We can write $u_n$ and $V'(u_n)$ in terms of their fourier coefficients $z_{k,n}$ and $V'_{n,k}$ as

\begin{equation}
    u_n(t)=\sum_{k=-k_m}^{k_m}z_{k,n}\exp(i k \wb t); \quad V'(u_n(t))=
    \sum_{k=-k_m}^{k_m}V'_{k,n}\exp(i k \wb t),
\end{equation}

As $z_{k,n\pm 1}=z_{k,n}\,\exp(\pm i\alpha)$, substituting in Eq.~(\ref{eq:app:dyn}) we obtain

\begin{equation}
    \label{eq:app:fou1}
    -k^2\,\wb^2\,z_{k,n}+V'_{k,n}+4\ee\,z_{k,n}\sin^2({\alpha}/{2})\,.
\end{equation}

Let us calculate the coefficients $V'_{k,n}$ for the generic
$\phi^4$ potential $V(u_n)=\frac{1}{2}u_n^2+ \frac{1}{4}s u_n^4 $. Then $V'(u_n)=u_n+s u_n^3 $ or

\begin{equation}
    V'(u_n)=\sum_{k=-k_m}^{k_m}z_{k,n}\exp(i k \wb t)+s\Big(\sum_{k=-k_m}^{k_m}z_{k,n}\exp(i k \wb t )\Big)^3\, .
\end{equation}

As $u_n(t)$ is real $z_{1,n}=z_{-1,n}^*$ and, therefore, within the RWA $u_n(t)=z_{1,n}\exp(i \wb t)+z_{1,n}^* \exp(-i \wb t)=2|z_{1,n}|\cos(\wb t+\phi_{1,n})$ with $\phi_{1,n}=\arg(z_{1,n})$. For a given $n=p$, we can always set $\phi_{1,p}=0$ by choosing the appropriate time origin; that is $z_{1,p}\in\mathbb{R}$ and $z_{1,p}=z_{-1,p}$. We only need to find the coefficients for the site $n=p$ as we can deduce any other one by using $z_{k,n+p}=z_{k,p}\,\exp(i n \alpha)$. Let us write $u(t)\equiv u_p(t)$, $z_k\equiv z_{k,p}$ and so on, for simplicity. Then

\begin{eqnarray}
    &\mbox{}& V'(u(t))=(z_1 \exp(i \wb t)+ z_1\exp(-i \wb t)) +s (z_1\exp(i \wb t)+ z_1\exp(-i \wb t))^3= \nonumber \\
    &\mbox{}&z_1 (\exp(i \wb t)+ \exp(-i \wb t)) +s z_1^3 (\exp(i \wb t)+ \exp(-i \wb t))^3= \nonumber \\
    &\mbox{}&z_1 (\exp(i \wb t)+ \exp(-i \wb t)) + s z_1^3 (3 [\exp(i\wb t)+\exp(-i\wb t)] + [\exp(3 i\wb t)+\exp(-3 i\wb t)])=
    \nonumber\\
    &\mbox{}& (z_1 +3s\,z_1^2)(\exp(i \wb t)+ \exp(-i \wb t)) + s z_1^3( \exp(3 i\wb t)+\exp(-3 i\wb t))\, . \nonumber
\end{eqnarray}

Therefore, the coefficient of $\exp(i \wb t)$ is $V'_1=(z_1+3s\,z_1^2)$. Substitution in Eq.~(\ref{eq:app:fou1}) for $n=p$ (omitted) gives $-\wb^2 z_1+(z_1+3 s\,z_1^3)+4\ee z_1\sin^2(\alpha/2)=0\, $. This equation has two solutions $z_1=0$ and $z_1=(-\wb^2+1+4\ee\sin^2(\alpha/2))$, which becomes also zero for the critical value of the coupling parameter $\ee$ given by:

\begin{equation}\label{eq:critphi4}
    \ee_c=\frac{\wb^2-1}{4\sin^2(\alpha/2)}\,. \qquad (\phi^4 \textrm{potential})
\end{equation}

Therefore, $z_{1,p}\equiv z_1=0$ and so are $z_{1,n}$ and $u_n$
within the RWA. The flux $J_0$ also becomes zero, according to
Eq.~(\ref{eq:flux}). This result does not depends on the value of
$s$, which measures de degree of nonlinearity of the potential. As
any hard symmetric potential can be approximated by the $\phi^4$
potential with some $s>0$, $\ee_c$  is a good approximation for
all of them. However, for a soft symmetric potential $s$ is
negative, $\wb<1$, and as $\ee>0$ there is no value of the
coupling that nullifies $z_1$ and the flux cannot be nullified.

For the Morse potential, the RWA approximations is not so useful.
From the numerical results we know that the critical coupling
$\ee$ corresponds to a divergence of the flux $J_0$, and therefore
to a divergence of $u_n$ and its first fourier coefficient
$z_{1,n}$. Then, $V'(u_n)=2(\exp(-2u_n)-\exp(-u_n))$ tends to zero
when $u_n\rightarrow\infty$ and so does $V'_{1,n}$. By
substitution in Eq.~(\ref{eq:app:fou1}) for $k=1$, neglecting
$V'_{1,n}$ and simplifying  $z_{1,n}$ we get
$-\wb^2+4\ee\,\sin^2({\alpha}/{2})=0$ and the critical coupling is

\begin{equation}\label{eq:critMorse}
    \ee_c=\frac{\wb^2}{4\sin^2(\alpha/2)}\,. \qquad \textrm{(Morse potential)}\,
\end{equation}

We have not been able to deduce the reason why $u_n$ becomes infinity.

\end{document}